\documentclass[aps,prd,showpacs,nofootinbib,preprintnumbers]{revtex4}
\usepackage{epsfig}

\newcommand{\la}[1]{\label{#1}}
\newcommand{\be}{\begin{equation}}
\newcommand{\ee}{\end{equation}}
\newcommand{\ba}{\begin{eqnarray}}
\newcommand{\ea}{\end{eqnarray}}
\newcommand{\rmi}[1]{{\mbox{\scriptsize #1}}}
\newcommand{\fig}{Fig.~}
\newcommand{\eq}{Eq.~}
\newcommand{\se}{Sec.~}
\newcommand{\eqs}{Eqs.~}
\newcommand{\nr}[1]{(\ref{#1})}

\newcommand{\nn}{\nonumber \\}
\newcommand{\fr}[2]{{\frac{#1}{#2}\,}}
\newcommand{\msbar}{{\overline{\mbox{\rm MS}}}}

\renewcommand{\vec}[1]{{\bf #1}}

\newcommand{\bG}{{\beta_G}}

\newcommand{\tinymsbar}{{\overline{\mbox{\tiny\rm{MS}}}}}
\newcommand{\Lambdamsbar}{{\Lambda_\tinymsbar}}
\newcommand{\Nf}{N_{\rm f}}
\newcommand{\nF}{n_{\rm F}}
\newcommand{\nB}{n_{\rm B}}
\newcommand{\nFm}[1]{\hat \nF (\sqrt{#1 + y} - \hat\mu)}
\newcommand{\nFp}[1]{\hat \nF (\sqrt{#1 + y} + \hat\mu)}
\newcommand{\Nc}{N_{\rm c}}
\newcommand{\Tc}{T_{\rm c}}
\newcommand{\rmO}{{\mathcal{O}}}
\newcommand{\bmu}{\bar\mu}
\newcommand{\boldmu}{\mbox{\boldmath$\mu$}}

\newcommand{\bM}{\beta_\rmi{M}}
\renewcommand{\bG}{\beta_\rmi{G}}
\newcommand{\aE}[1]{\alpha_\rmi{E#1}}
\newcommand{\aEms}[1]{\alpha_\rmi{E#1}^\tinymsbar}
\newcommand{\bE}[1]{\beta_\rmi{E#1}}
\newcommand{\bEms}[1]{\beta_\rmi{E#1}^\tinymsbar}

\def\lsi{\raise0.3ex\hbox{$<$\kern-0.75em\raise-1.1ex\hbox{$\sim$}}}
\def\gsi{\raise0.3ex\hbox{$>$\kern-0.75em\raise-1.1ex\hbox{$\sim$}}}

\newcommand{\gsim}{\mathop{\gsi}}

\newcommand{\Tint}[1]{{\hbox{$\sum$}\!\!\!\!\!\!\!\int}_{\!\!\!\!#1}}

\newcommand{\mi}{m_i} % m_{Bi}
\newcommand{\g}{g} % g_B

\begin{document}

\title{Quark mass thresholds in QCD thermodynamics}
\author{Mikko \surname{Laine}}
\email{laine@physik.uni-bielefeld.de}
\affiliation{Faculty of Physics, University of Bielefeld, 
D-33501 Bielefeld, Germany}
\author{York \surname{Schr\"oder}}
\email{yorks@physik.uni-bielefeld.de}
\affiliation{Faculty of Physics, University of Bielefeld, 
D-33501 Bielefeld, Germany}

\begin{abstract}
We discuss radiative corrections to how quark mass 
thresholds are crossed, as a function of the temperature, 
in basic thermodynamic observables such as the pressure, the energy 
and entropy densities, and the heat capacity of high temperature QCD. 
The indication from leading order that the charm quark 
plays a visible role at surprisingly low temperatures, is confirmed. 
We also sketch a way to obtain phenomenological estimates relevant 
for generic expansion rate computations at temperatures between 
the QCD and electroweak scales, pointing out where improvements 
over the current knowledge are particularly welcome. 
\end{abstract}
\preprint{BI-TP 2006/07, hep-ph/0603048}
\maketitle

\section{Introduction}

Besides being of fundamental theoretical 
interest to finite temperature field theory, 
the thermodynamic pressure of the Standard Model, 
as a function of the temperature $T$ and of various chemical 
potentials $\mu_i$, has several potential phenomenological applications. 
Most notably it dictates, through the Einstein equations, 
the expansion rate of the radiation dominated Early Universe. 
The expansion rate
in turn determines when various dark matter candidates decouple, 
thus fixing their relic densities: fine details of the pressure 
could become observable for instance if dark matter is made  
of electroweak scale WIMPs~\cite{olddm2,dm2} or
of keV-scale sterile neutrinos~\cite{olddm1,dm1}.
Furthermore, the pressure is in principle visible 
in the present-day  
spectrum of the gravitational wave background that 
was generated during the inflationary epoch~\cite{djs}. 
More generally, the pressure incorporates the fact 
that the Standard Model possesses a trace anomaly, 
i.e. ${T^\mu}_\mu \neq 0$, which in turn can 
influence many kinds of gravity-related cosmological scenarios
(for recent examples, see Refs.~\cite{xxx}).

Apart from cosmology, the pressure is 
potentially also relevant for the hydrodynamic 
expansion that the dense matter generated in current and upcoming
heavy ion collision experiments may undergo. In this case 
there is some room for caution, however, since the issue
of whether local thermodynamic equilibrium is reached 
remains controversial~\cite{urhic}. 

Given that the
biggest theoretical challenges are related to
strongly interacting particles, considerable
efforts have been devoted to the determination
of the QCD part of the pressure over a course
of years. Denoting by $g$ the renormalised strong coupling constant, 
perturbative corrections to the non-interacting Stefan-Boltzmann
law have been determined at relative orders 
$\rmO(g^2)$~\cite{es}, 
$\rmO(g^3)$~\cite{jk}, 
$\rmO(g^4\ln(1/g))$~\cite{tt}, 
$\rmO(g^4)$~\cite{az}, 
$\rmO(g^5)$~\cite{zk}, and 
$\rmO(g^6\ln(1/g))$~\cite{gsixg}, 
as a function of the number of colours, $\Nc$, and the number
of massless quark flavours, $\Nf$. 
The first presently unknown order, $\rmO(g^6)$, contains non-perturbative 
coefficients~\cite{linde,gpy}, but those can also be 
attacked~\cite{plaq,nonpert}.
All orders of $g$ are available in the formal limit of large $\Nf$~\cite{Nf}.
These results have been extended to the case of finite
quark chemical potentials~\cite{av1,av2,Nfmu}, and a similar 
computation has recently also been 
finalised for the weakly interacting part of the Standard
Model, at temperatures higher than the electroweak scale~\cite{gv}. 
Moreover, the fact that several orders are available allows 
to experiment with various kinds of resummations~\cite{bir,pade}.

Surprisingly, however, relatively little seems to be known 
about the dependence of the QCD pressure on the quark masses $m_i$, 
$i = 1, ..., \Nf$. While the non-interacting
Stefan-Boltzmann law is 
readily extended to this situation, it in fact appears
that even the first non-trivial term, $\rmO(g^2)$, has not
been exhaustively investigated in the literature
(see, however, Ref.~\cite{xj}). In principle
this term has of course been available since almost 30 years~\cite{jk}, 
but in explicit form 
only in a renormalization scheme for quark masses which  
differs from the current standard, the $\msbar$ scheme.
Furthermore, no general numerical evaluation of the basic
integrals appearing has been presented, as far as we know. 
For $T=0$ but $\mu_i\neq 0$, the full $\rmO(g^2)$ analysis 
has also only been carried out recently~\cite{fr}.

Several probable reasons for the apparent lack of interest can 
surely be envisaged. First of all, the dependence on $\Nf$
is known to a high order in the massless case, 
and interpolating between integer 
values of $\Nf$ should give much of the information that we may 
need for the massive case. 
Second, including quark masses turns out to be technically
cumbersome~\cite{jk}. Third, 
there are several indications, for instance from considerations
of the baryon chemical potential~\cite{av1,av2}
and of mesonic correlation lengths~\cite{lv}, that the 
convergence is much better in the quark sector than in 
the gluonic sector, so that the lowest non-trivial order may
already provide sufficient accuracy. Nevertheless, we feel that 
the last assumption deserves to be checked, at least at 
the next-to-leading order (NLO) $\rmO(g^2)$, and this 
is the purpose of the present paper.  

In short, our general philosophy will then be to account for 
the gluonic contributions to the highest
order available, $\rmO(g^6\ln(1/g))$, and consider  
the change that quarks with finite physical masses inflict 
on this result at NLO, $\rmO(g^2)$. We do find that the 
quark mass effects at NLO are not too different from those
at the leading order, $\rmO(g^0)$,
such that the philosophy of terminating
at $\rmO(g^2)$ is at least self-consistent. Nevertheless, we
also outline the procedure for determining the quark mass
dependence up to the order $\rmO(g^6)$.

Apart from the theoretical goals mentioned, we also wish to 
sketch certain phenomenological results in this paper. Consider 
the temperature evolution of an expanding system in the case of
cosmology, for instance. Einstein equations then lead to
(see, e.g., Ref.~\cite{ikkl})
\be
 \frac{1}{T} \frac{{\rm d}T}{{\rm d}t} = 
 - \sqrt{\frac{24\pi e(T)}{m_\rmi{Pl}^2}}
 \frac{s(T)}{c(T)}
 \;, \la{dTdt}
\ee
where $t$ is the time; 
we assumed the Universe to be flat ($k=0$); 
and we ignored the cosmological constant.
All the quantities appearing here follow from the pressure: 
$s(T) = p'(T)$ is the entropy density,  
$e(T) = T s(T) - p(T)$ is the energy density, 
and 
$c(T) = e'(T) = T p''(T)$ is the heat capacity.
We wish to present our favoured ``fits'' for all these
functions for temperatures between the QCD 
and electroweak scales, indicating where
further  work is required.

The plan of this paper is the following. 
We start by elaborating on the basic formalism in~\se\ref{se:basic}, 
discuss quark mass thresholds in~\se\ref{se:masses},
present a phenomenological evaluation of the various 
thermodynamic functions relevant for physical QCD in~\se\ref{se:phenqcd}, 
include weakly interacting particles in~\se\ref{se:phenew}, 
and conclude in~\se\ref{se:concl}.

%%%%%%%%%%%%%%%%%%%%%%%%%%% SECTION %%%%%%%%%%%%%%%%%%%%%%%%%%%%%%%%%%%%%
%
\section{Basic formalism}
\la{se:basic}

In order to determine the basic thermodynamic
quantities of the Standard Model, all of which 
can be derived from minus the grand canonical free energy density, 
or the pressure $p(T,\boldmu)$, where
$\boldmu$ collects together the various chemical 
potentials associated with conserved global 
charges,\footnote{%
  The notation $p(T,\boldmu)$ always 
  implicitly refers to the ultraviolet finite 
  difference $p(T,\boldmu) - p(0,\mathbf{0})$. 
  }
we make use of the framework of dimensionally 
reduced effective field theories~\cite{dr,generic,bn}.
This framework allows to organise the computation in 
a transparent way, and implements various resummations 
of higher order effects. 
We start by briefly reviewing certain aspects of the general 
framework for QCD; 
further details can be found in Ref.~\cite{gsixg}.

Dimensional reduction proceeds by first integrating out the ``hard 
modes'', with momenta or Matsubara frequencies of order $2\pi T$.
This produces an effective theory~\cite{dr}, called EQCD~\cite{bn}, 
which is a three-dimensional SU($\Nc$) gauge theory with 
a scalar field in the adjoint representation. The effective 
theory has a certain number of couplings, parametrised by 
functions denoted by $\aE{1}...\aE{7}$ [contributing
up to $\rmO(g^6\ln(1/g))$] and $\bE{1}...\bE{5}$
[contributing at $\rmO(g^6)$] in Ref.~\cite{gsixg}; 
in the following we explicitly specify the definitions
for only a subset of them. These parameters contain all 
the information concerning the hard modes. 
Assuming the use of dimensional
regularization, we denote by $\aEms{i}$, $\bEms{i}$ parameters from which 
the $1/\epsilon$-divergences have been removed by the $\msbar$-prescription.

To proceed, we need to specify explicitly 
the effective mass parameter $m_3^2$
and the effective gauge coupling $g_3^2$ of EQCD at NLO: 
\ba
 \hat m_3^2 \; \equiv \;  
 \frac{m_3^2}{T^2} & \equiv & 
 g^2 \aEms{4} + \frac{g^4}{(4\pi)^2} \aEms{6}
 \;, \la{m32} \\ 
 \hat g_3^2 \; \equiv \; 
 \frac{g_3^2}{T} & \equiv & 
 g^2 +  \frac{g^4}{(4\pi)^2} \aEms{7}
 \;. \la{g32}
\ea
Both parameters are renormalization group invariant up to the 
order computed, i.e., the dependence on the scale parameter 
$\bmu$ is of order ${\cal O}(g^6)$.

With this notation, the physical pressure of hot QCD can 
be written in the form
\be
 p_\rmi{QCD} \equiv p_\rmi{hard} + p_\rmi{soft} 
 \;, \la{sum}
\ee
where $p_\rmi{hard}$ represents the contribution of the hard 
modes (by definition containing both all direct hard contributions 
to the pressure, and all finite terms emerging from products 
like $\epsilon \cdot 1/\epsilon$), while $p_\rmi{soft}$ represents 
the contribution of the soft modes.
Up to the accuracy $\rmO(g^6)$,
$p_\rmi{hard}$ can conveniently be expressed as
\ba
 \frac{p_\rmi{hard}}{T^4} & = & \aEms{1}
 + \hat g_3^2 \aEms{2}
 + \frac{\hat g_3^4}{(4\pi)^2} 
 \Bigl(\aEms{3} - \aEms{2}\aEms{7} - \fr14 d_A C_A \aEms{5} \Bigr) 
 + \nonumber \\[1mm]  & + & 
 \frac{\hat g_3^6}{(4\pi)^4} 
 \biggl\{ 
 \Bigl[
 d_A C_A (\aEms{6} - \aEms{4} \aEms{7}) 
 - d_A C_A^3 \Bigl( \frac{43}{3} - \frac{27}{32}\pi^2 \Bigr)
 \Bigr] \ln\frac{\bmu}{4\pi T} 
 + \Delta_\rmi{hard}
 \biggr\} 
 \;, \hspace*{0.9cm}
 % \nn & &  
 \la{phard2}
\ea
where $d_A \equiv \Nc^2-1, C_A \equiv \Nc$, and we have 
separated a term on the last line which cancels the 
$\bmu$-dependence of $p_\rmi{soft}$ at $\rmO(g^6)$.
The function $\Delta_\rmi{hard}$, 
\ba
 \Delta_\rmi{hard} & \equiv & 
  \Bigl[
  d_A C_A (\aEms{6} - \aEms{4} \aEms{7}) 
 - d_A C_A^3 \Bigl( \frac{43}{3} - \frac{27}{32}\pi^2 \Bigr)
 \Bigr] \ln\frac{4\pi T}{\bmu} 
 + \nonumber \\[1mm] & & + 
 \bEms{1} + 2 \aEms{2} (\aEms{7})^2 - 2 \aEms{3} \aEms{7}
 -\fr14 d_A C_A \Bigl( 
 \bEms{2} - \aEms{5} \aEms{7} + \aEms{4} \bEms{3}
 \Bigr)
 \;, \hspace*{0.9cm}
\ea 
depends on $\Nc,\Nf, m_i, \mu_i$, and $\bmu/T$. 
The contributions of the soft modes are~\cite{gsixg}
\ba
 \frac{p_\rmi{soft}}{T^4} & = & 
 \frac{\hat m_3^3}{12\pi}  d_A
 - \frac{\hat g_3^2 \hat m_3^2}{(4\pi)^2} d_AC_A
 \biggl( 
   \ln\frac{\bmu}{2 m_3} + \fr34
 \biggr) -
 % \nn & & - 
 \frac{\hat g_3^4 \hat m_3}{(4\pi)^3} d_A C_A^2
 \biggl( 
   \frac{89}{24} + \frac{\pi^2}{6} - \frac{11}{6} \ln 2
 \biggr)
 + \hspace*{6mm}
 \nonumber \\[1mm] & + & 
  \frac{\hat g_3^6}{(4\pi)^4} d_A C_A^3
  \biggl[
  \biggl(
   \frac{43}{4} - \frac{491}{768} \pi^2 
  \biggr) \ln\frac{\bmu}{2 m_3} 
 + 
  \biggl(
   \frac{43}{12} - \frac{157}{768} \pi^2 
  \biggr) \ln\frac{\bmu}{2 C_A g_3^2} 
  + \Delta_\rmi{soft} 
  \biggr]
 % + \nn & & 
 % + \rmO\Bigl(\frac{g_3^8}{m_3}\Bigr)
 \;. 
%  \nn & & 
 \la{psoft}
\ea
The function $\Delta_\rmi{soft}$ reads
\be
 \Delta_\rmi{soft} =
 % [\mbox{numerical value of }]
 \bM + \bG - \aEms{4}
 \biggl[ \frac{d_A + 2}{4C_A^3} \bEms{4} + 
         \frac{2 d_A -1}{4C_A^4} \bEms{5} \biggr]
 \;, \la{Dsoft}
\ee
where $\bM$ can be found in Ref.~\cite{aminusb}, 
and a numerical estimate of $\bG$ in Ref.~\cite{nonpert}. 

Let us stress that the formulae presented
apply independently of whether quark masses are included or not:
all quark mass effects can be incorporated in the perturbative
functions $\aEms{1}...\aEms{7}$, $\bEms{1}...\bEms{5}$. 
In particular, the non-perturbative
numerical value $\bG$ and the contribution
from the Debye scale $\bM$ in \eq\nr{Dsoft} are ``universal''.

In the following, we refer to the various orders of 
the weak-coupling expansion  according to 
the power of $\hat g_3, \hat m_3$ that appear, with 
the rule $\rmO(\hat m_3) = \rmO(\hat g_3) = \rmO(g)$. In other words, 
``$\rmO(g^n)$'' denotes $\rmO(\hat g_3^{n-k} \hat m_3^k)$ 
in the expression constituted by \eqs\nr{sum}, \nr{phard2}, \nr{psoft}.
If $\hat g_3^2, \hat m_3^2$ were to be re-expanded 
in terms of $g^2$, the result of Ref.~\cite{gsixg}
would be reproduced up to $\rmO(g^6)$. In practice, however, 
it is advisable to keep the result in an unexpanded form, because 
this makes it more manageable, and because the unexpanded form
introduces resummations of higher order contributions.

%%%%%%%%%%%%%%%%%%%%%%%%%%% SECTION %%%%%%%%%%%%%%%%%%%%%%%%%%%%%%%%%%%%%
%
\section{Quark mass thresholds in the pressure}
\la{se:masses}

In the absence of an explicit $\rmO(g^6\ln(1/g))$-computation
with $\mi\neq 0$, we are forced to estimate the effects of finite
quark masses by other means. In order to do this, we take as the 
starting point the limit $\Nf = 0$, where all quarks are treated
as infinitely heavy. Subsequently, the masses of a number $\Nf$ 
of them are lowered to their physical values. In this process
the pressure at any given temperature increases. The goal is then 
to estimate the ``correction factors'' that describe this increase.   
Note that this philosophy corresponds to the ``unquenching'' of 
quark effects that is regularly attempted in lattice simulations.

The philosophy just suggested is of course not unique. 
For instance, one could also start from the ``opposite'' limit, 
taking as the starting point the pressure for a fixed $\Nf$, 
but with vanishing quark masses, and then increase the masses. 
In fact, one could perhaps even define an effective non-integer 
massless $\Nf$ by evaluating the massive $\aEms{1}$ (\eq\nr{aE1} below), 
fitting it to the massless formula (\eq(A.1) in Ref.~\cite{gsixg}),
and using the resulting $\Nf$ in the
massless result of $\rmO(g^6\ln(1/g))$~\cite{kr}.
We prefer, however, to take here the infinitely massive case $\Nf = 0$ 
as the reference point, since this limit can be treated with more
confidence than the chiral limit, as we describe in the next section, 
and since this limit offers us a convenient way to probe the 
convergence of our recipe, as we now explain.  

The simplest thinkable way to estimate the correction 
factors due to non-infinite quark masses would be to multiply 
the $\Nf = 0$ result with the change indicated by the 
Stefan-Boltzmann law, i.e. by $\aEms{1}(\Nf)/\aEms{1}(0)$~\cite{dm2}.
The first non-trivial improvement of this philosophy is then  
to determine the functions $\aEms{1}, \aEms{2}, \aEms{7}$ 
in the general massive case, which allows us to evaluate
the order $\rmO(g^2)$ result for the pressure, {\it viz.} 
\be 
 \frac{p_\rmi{QCD}}{T^4} \approx 
 \aEms{1} + \hat g_3^2 \aEms{2}
 \;. 
\ee
Afterwards, we may modify the $\Nf = 0$ 
result with a correction factor,
\be
  \frac{
  [\aEms{1} + \hat g_3^2 \aEms{2}](\Nf)}{
  [\aEms{1} + \hat g_3^2 \aEms{2}](0)}
 \;. \la{corrfac}
\ee
Comparing the outcome of this $\rmO(g^2)$ recipe
with the corresponding $\rmO(g^0)$ recipe allows to probe the convergence. 
Note that it is important to also determine $\aEms{7}$, since only 
this way can the renormalisation scale that appears in $\hat g_3^2$ 
be reasonably fixed (cf.\ \eq\nr{g32}).

%%%%%%%%%%%%%%%%%%%%%%%%%%%%%%%%% FIGURE %%%%%%%%%%%%%%%%%%%%%%%%%%%%%%%%%
\begin{figure}[t]

\centerline{%
\epsfysize=6.0cm\epsfbox{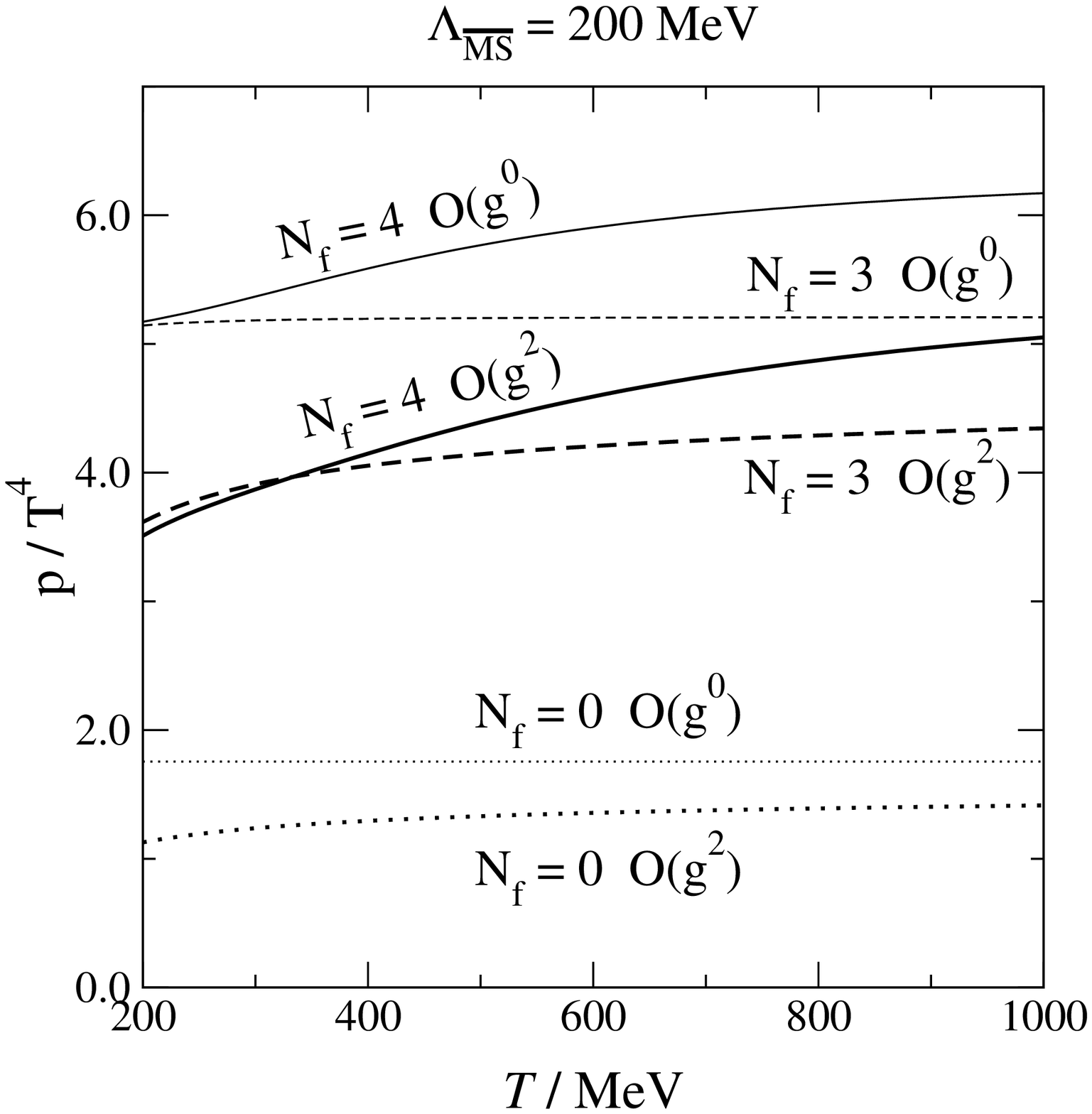}%
~~~~\epsfysize=6.0cm\epsfbox{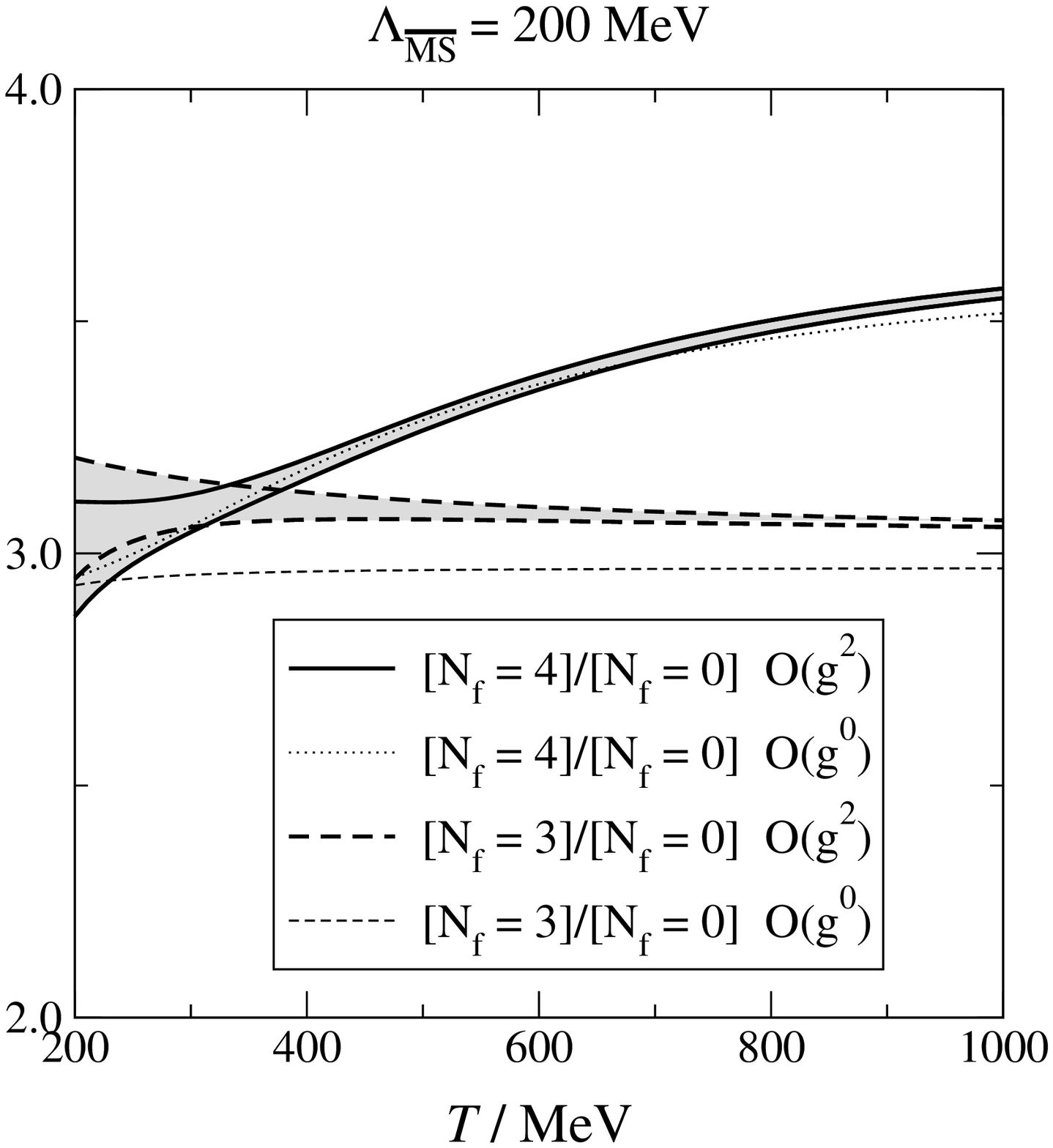}%
% ~~\epsfysize=5.0cm\epsfbox{}%
}

\caption[a]{%% \small 
Left: the pressure for $\Nf = 0$, 3, 4, 
at $\rmO(g^0)$ and $\rmO(g^2)$. Right: the ``correction factors'' accounting
for the effects of quarks, at $\rmO(g^0)$ and $\rmO(g^2)$
(cf.\ \eq\nr{corrfac}). 
They grey bands indicate the effect of $\msbar$ scheme scale 
variations by a factor \linebreak 0.5 ... 2.0 around the ``optimal'' value. 
It is observed that while the $\rmO(g^2)$ corrections 
are of order 20...30\%  in the pressure, they are of 
order 5\% in the correction factors for $\Nf = 3$, 
and even less for the physical case $\Nf = 4$. 
}

\la{fig:mass}
\end{figure}
%%%%%%%%%%%%%%%%%%%%%%%%%%%%%%%%%%%%%%%%%%%%%%%%%%%%%%%%%%%%%%%%%%%%%%%%%%%

We thus proceed to compute $\aEms{1}$, $\aEms{2}$, $\aEms{7}$.
We do this in full generality, keeping $\Nf$, $\Nc$, the quark 
masses $m_i$, and the chemical potentials $\mu_i$ as free parameters. 
The quark masses and the strong gauge coupling are renormalised
in the $\msbar$ scheme. Some details concerning the computation
are collected in Appendix A. As final results, we obtain 
\ba
 \aEms{1}  \!\! & = & \!\!  
 d_A \frac{\pi^2}{45} + 4 C_A 
 \sum_{i=1}^{\Nf} F_1\biggl( \frac{m_i^2}{T^2},\frac{\mu_i}{T} \biggr)
 \;, \la{aE1} \\[1mm]
 \aEms{2} \!\! & = & \!\!
 -\frac{d_AC_A}{144} - d_A \sum_{i=1}^{\Nf}
 \biggl\{ 
 \fr16 
   F_2\biggl( \frac{m_i^2}{T^2},\frac{\mu_i}{T} \biggr)
    \biggl[ 1 + 6 
   F_2\biggl( \frac{m_i^2}{T^2},\frac{\mu_i}{T} \biggr) \biggr]
 + \nonumber \\[1mm] & & + 
 \frac{m_i^2}{4\pi^2 T^2} \biggl( 
 3 \ln \frac{\bmu}{m_i} + 2
 \biggr)
 F_2\biggl( \frac{m_i^2}{T^2},\frac{\mu_i}{T} \biggr)
 - \frac{2 m_i^2}{T^2} 
 F_4\biggl( \frac{m_i^2}{T^2},\frac{\mu_i}{T} \biggr)
 \biggr\}
 \;, \la{aE2} \\[1mm]
 \aEms{7} \!\! & = & \!\!
 \frac{22C_A}{3}
 \biggl[
   \ln\biggl( \frac{\bmu e^{\gamma_E}}{4\pi T} 
   \biggr) + \frac{1}{22}
 \biggr] 
 -  % \nn & & - 
 \fr23 \sum_{i=1}^{\Nf}
 \biggl[
  2 \ln \frac{\bmu}{m_i} + 
  F_3\biggl( \frac{m_i^2}{T^2},\frac{\mu_i}{T} \biggr)
  \biggr]
 \;, \la{aE7}
\ea
where the functions $F_1,...,F_4$ and some of their properties
are detailed in Appendix B.

To estimate the numerical importance of the $\rmO(g^2)$
corrections, we need to assign a value to all the parameters
that appear. Following a simple-minded logic, we use 1-loop
running,  
\be
 g^2(\bmu) = \frac{24\pi^2}{(11 C_A - 4 T_F) \ln(\bmu/\Lambdamsbar)}
 \;, \quad
 m_i(\bmu) = m_i(\bmu_\rmi{ref})
 \biggl[
   \frac{\ln(\bmu_\rmi{ref}/\Lambdamsbar)}{\ln(\bmu/\Lambdamsbar)} 
 \biggr]^{\frac{9 C_F}{11C_A - 4 T_F}}
 \;, 
\ee
where $T_F = \Nf/2$, $C_F = (\Nc^2-1)/2 \Nc$,  
$\bmu_\rmi{ref} \equiv 2$~GeV. The quark masses 
at $\bmu = \bmu_\rmi{ref}$ are taken from Ref.~\cite{pdg}.
To choose $\bmu$, we apply the principle of minimal sensitivity 
criterion for the parameter $\hat g_3^2$, as suggested 
in Ref.~\cite{adjoint}. Furthermore, for illustration, 
we set $\Lambdamsbar \equiv 200$~MeV.

The outcome of this procedure is shown in \fig\ref{fig:mass}, 
for $\mu_i = 0$.
It is observed that while the $\rmO(g^2)$ corrections 
are of order 20...30\%  in the pressure (left panel), 
the ``correction factors'', i.e. the ratios in \eq\nr{corrfac},
only contain $\rmO(g^2)$ corrections of order 5\% for $\Nf = 3$, 
and even less for the physical case $\Nf = 4$ (right panel). 
This implies that the quark mass dependence of the pressure
probably converges  faster than the weak-coupling expansion as a whole. 

Finally, we note from \fig\ref{fig:mass}(right) that the charm
quark contribution starts to be visible already at fairly low
temperatures. At leading order, the quark mass dependence
is determined by the function $F_1$ (cf.\ \eq\nr{aE1}), 
which at low temperatures has the familiar classical form 
\be
 F_1\Bigl( \frac{m^2}{T^2} , 0\Bigr)
 \approx \biggl( \frac{m}{2\pi T} \biggr)^\fr32 
 \exp\Bigl( -\frac{m}{T} \Bigr) 
 \;. \la{F1appro}
\ee
It is observed that $F_1$ obtains 
5\% of its asymptotic value $7\pi^2/720$ at temperatures
as low as $T \approx m/5$. (For the precise numerical 
values of $F_1$, see \fig\ref{fig:Fis}.) As \fig\ref{fig:mass}(right)
shows, the onset of a visible charm mass dependence is postponed to 
about $T \sim 350$~MeV at $\rmO(g^2)$, but the basic pattern 
remains unchanged.

%%%%%%%%%%%%%%%%%%%%%%%%%%% SECTION %%%%%%%%%%%%%%%%%%%%%%%%%%%%%%%%%%%%%
%
\section{Phenomenological results for QCD}
\la{se:phenqcd}

We now move from fairly well-defined expressions towards phenomenology. 
The goal is to present, where possible, an educated numerical guess
for the physical QCD pressure. We set all chemical potentials to 
zero in the following.

The general philosophy we adopt is that, for temperatures above
the deconfinement transition, the weak-coupling expansion needs
to be evaluated up to the order where the dominant contributions
from all the different scales ($2\pi T$, $gT$, $g^2T$)
have made their entrances. There
is some support for such a recipe from a number of non-trivial 
observables~\cite{hp,gE2}. In practice, this means that the QCD
pressure would need to be evaluated up to $\rmO(g^6)$. 

Unfortunately, one of the $\rmO(g^6)$ terms, parametrised by 
$\bEms{1}$ in \se\ref{se:basic}, remains completely unknown for 
the moment even in the massless limit: it is a function of $\Nf$ 
and requires a perturbative 4-loop computation. This introduces
a certain unknown ``constant'' into the prediction. We propose
to fix the constant by the following recipe.

Let us start by considering the case $\Nf = 0$, $\Nc = 3$. 
Then the expressions in \se\ref{se:basic}
depend on only two parameters: 
on $\bmu/T$ (through $\aEms{1}...\aEms{7}$, $\bEms{1}...\bEms{5}$), 
and on $\bmu/\Lambdamsbar$ (through $g^2(\bmu)$). 
It so happens that the dependence on $\bmu$, 
which formally cancels up to the order of the computation, 
is numerically non-monotonous
(see, e.g., Ref.~\cite{bir}), so that the specific choice is not 
terribly important, as long as we are close to the extremum. 
In practice we choose $\bmu/T$ according to the 
principle of minimal sensitivity criterion for the 
parameter $\hat g_3^2$, as already mentioned. 
Thereby the results only depend on $T/\Lambdamsbar$ 
and on the unknown $\rmO(g^6)$ terms, contained 
in $\Delta_\rmi{hard}$ and $\Delta_\rmi{soft}$, 
defined in \eqs\nr{phard2}, \nr{psoft}.
It is important to note that once $\bmu/T$ has been fixed, 
the $\Delta$'s can be treated as temperature-independent
constants. It is furthermore
convenient to combine them into a single term,\footnote{%
  We note that $\Delta$ differs by a certain constant from 
  a similar constant employed in the figures of Ref.~\cite{gsixg}.
  } 
\be 
 \Delta_\rmi{hard} + 
 d_A C_A^3 
 \Delta_\rmi{soft} 
 \equiv 
 d_A C_A^3  \Delta
 \;.
\ee

In order to now eliminate the dependence on $\Delta$, 
we ``match'' the perturbative prediction to 4d lattice simulation
results for the case $\Nf = 0$, where the continuum limit has 
been reached with reasonable precision~\cite{bi1,Nf0}. 
It should be stressed that this step is purely phenomenological: 
in principle $\Delta$ is computable from the theory. 
On the other hand, there is every reason to expect that results
obtained through the dimensionally reduced framework do match
4d lattice results as soon as $T \gsim 2 \Tc$, where $\Tc$ is 
the temperature of the deconfinement phase transition 
(see, e.g., Refs.~\cite{hp}, \cite{mu}--\cite{bmp}). 
Moreover, that a family of 
functions specified by a single parameter should match 
a given function for a whole range of argument values, 
provides for a non-trivial consistency check. 

%%%%%%%%%%%%%%%%%%%%%%%%%%%%%%%%% FIGURE %%%%%%%%%%%%%%%%%%%%%%%%%%%%%%%%%
\begin{figure}[t]

\centerline{%
\epsfysize=6.0cm\epsfbox{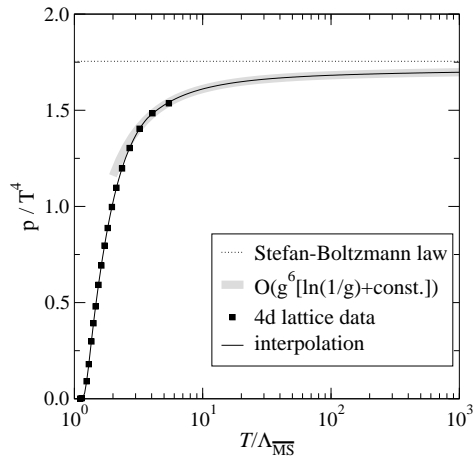}%
% ~~\epsfysize=5.0cm\epsfbox{}%
% ~~\epsfysize=5.0cm\epsfbox{}%
}

\caption[a]{%% \small 
A phenomenological interpolating curve (solid line) 
for the QCD pressure at $\Nf = 0$. In the perturbative curve (grey band) 
the unknown $\rmO(g^6)$ constant has been adjusted so that lattice data
(closed squares~\cite{bi1}) is matched, once $T > 3.6 \Lambdamsbar$.}

\la{fig:glueinter}
\end{figure}
%%%%%%%%%%%%%%%%%%%%%%%%%%%%%%%%%%%%%%%%%%%%%%%%%%%%%%%%%%%%%%%%%%%%%%%%%%%

Now, lattice results are usually presented in terms of $T/\Tc$, 
rather than $T/\Lambdamsbar$. We thus need a value for 
$\Tc/\Lambdamsbar$; we use $\Tc/\Lambdamsbar \approx 1.20$
which appears to be consistent with all independent 
determinations (cf.\ Ref.~\cite{gE2}, Sec.~4.2). After this choice, 
an excellent match can be obtained (we do this by minimising
the difference squared of the function values in the range
$T > 3 \Tc$), with a value
$\Delta \approx -3.287$ (cf.\ \fig\ref{fig:glueinter}).
In the following we will take the cubic spline 
interpolation shown in \fig\ref{fig:glueinter} as the ``starting point'', 
which will then be ``corrected'' by the effects of quarks. 

To now include quarks, we simply multiply
the result just obtained by the correction factor in \eq\nr{corrfac}.
We should expect this construction to work the better the higher 
the temperature, but surely at least $T > 200$ MeV is required. 

It needs to be noted, however, that like in \fig\ref{fig:mass},
the evaluation of the correction factor necessitates fixing 
$\Lambdamsbar$ in physical units. This exercise is non-trivial. 
We again choose a purely phenomenological but rather convenient 
procedure, which makes use of the 
pressure produced by the full set of hadronic resonances~\cite{pdg}.
Indeed, it has been demonstrated recently that if the resonance masses
are tuned to correspond to the quark masses accessible to current
lattice simulations, the resulting ``resonance gas pressure'' 
works surprisingly well even for temperatures deep into 
the crossover region~\cite{krt}.

Thus, we tune $\Lambdamsbar$ such that our analytic recipe 
and the first derivative thereof match the resonance gas result
(in which the temperature is automatically measured in physical units)
at a certain temperature. 
Examples are shown in \fig\ref{fig:qcdpoT4}(left). 
The values of $\Lambdamsbar$ that result depend slightly on $\Nf$ 
and on variations of quark masses within their experimental 
errors, but the typical range is 
$\Lambdamsbar \approx 175 ... 180$~MeV. 
We should stress that this matching is of course rather arbitrary, 
but it does produce qualitatively reasonable results with, 
for instance, an inflection point on the side $T > \Tc \sim 175$~MeV, 
as suggested by lattice results~\cite{Nfcp}--\cite{Nfbi}.

Naturally, the resonance gas results cannot really be trusted 
in quantitative detail for temperatures above, say, 150 MeV.
Therefore, for a certain interval (which we choose
to be $T = 150 ... 350$~MeV, and shade in all figures), 
the results remain to be established 
by lattice simulations. The matter becomes even more urgent, 
when one considers derivatives of the pressure, to which we now turn.

Apart from the pressure, its first and second derivatives
play an important role, as already mentioned in connection 
with~\eq\nr{dTdt}. There are various ways of presenting the 
information contained in these derivatives: we may for instance
parametrise the physical observables $e(T)$, $s(T)$, $c(T)$ 
through effective numbers of bosonic degrees of freedom, 
\ba
 g_\rmi{eff}(T)  \equiv  \frac{e(T)}{\Bigl[\frac{\pi^2 T^4}{30}\Bigr]}
 \;, \quad
 h_\rmi{eff}(T)  \equiv  \frac{s(T)}{\Bigl[\frac{2\pi^2 T^3}{45}\Bigr]}
 \;, \quad
 i_\rmi{eff}(T)  \equiv  \frac{c(T)}{\Bigl[\frac{2\pi^2 T^3}{15}\Bigr]}
 \;, \la{ieff} 
\ea
in terms of which \eq\nr{dTdt} becomes
\be
 \fr32 \sqrt{\frac{5}{\pi^3}}
 \frac{m_\rmi{Pl}}{T^3}
 \frac{{\rm d}T}{{\rm d}t} = - 
 \frac{\sqrt{g_\rmi{eff}(T)} h_\rmi{eff}(T)}{i_\rmi{eff}(T)}
 \;, 
\ee 
or we can consider dimensionless ratios like 
\ba
 w(T) & \equiv & \frac{p(T)}{e(T)} = \frac{p(T)}{Tp'(T) - p(T)}
 \;, \la{wT} \\ 
 c_s^2(T) & \equiv & \frac{p'(T)}{e'(T)} = \frac{p'(T)}{Tp''(T)}
 = \frac{s(T)}{c(T)}
 \;. \la{csT}
\ea
Both the ``equation-of-state'' $w(T)$ and the sound speed squared
$c_s^2(T)$ equal 1/3 in the non-interacting limit.
The deviation of the parameter $w(T)$ from 1/3 
is proportional to the trace anomaly, 
sometimes also called the interaction measure. 

%%%%%%%%%%%%%%%%%%%%%%%%%%%%%%%%% FIGURE %%%%%%%%%%%%%%%%%%%%%%%%%%%%%%%%%
\begin{figure}[t]

\centerline{%
\epsfysize=5.0cm\epsfbox{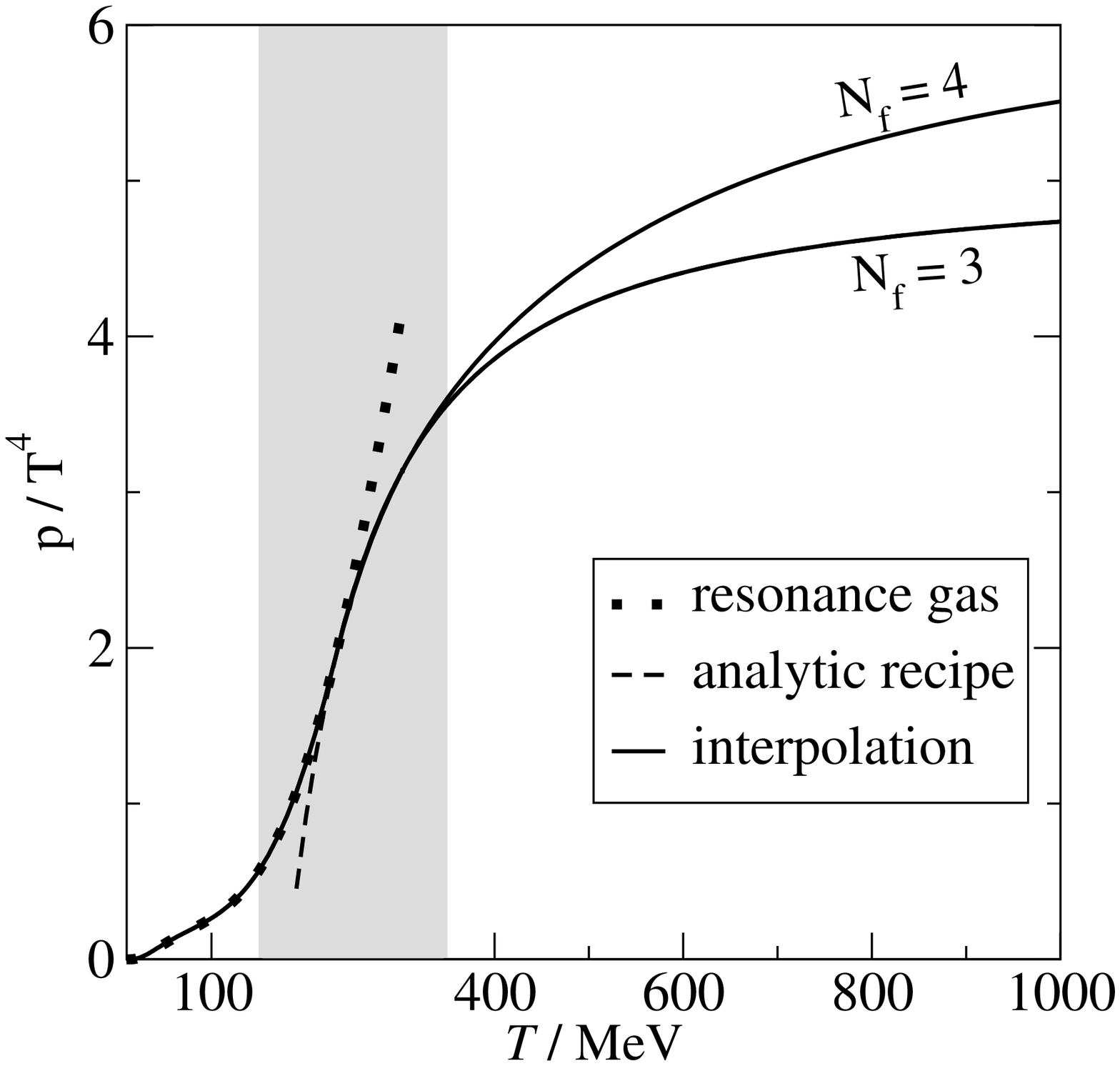}%
~~\epsfysize=5.0cm\epsfbox{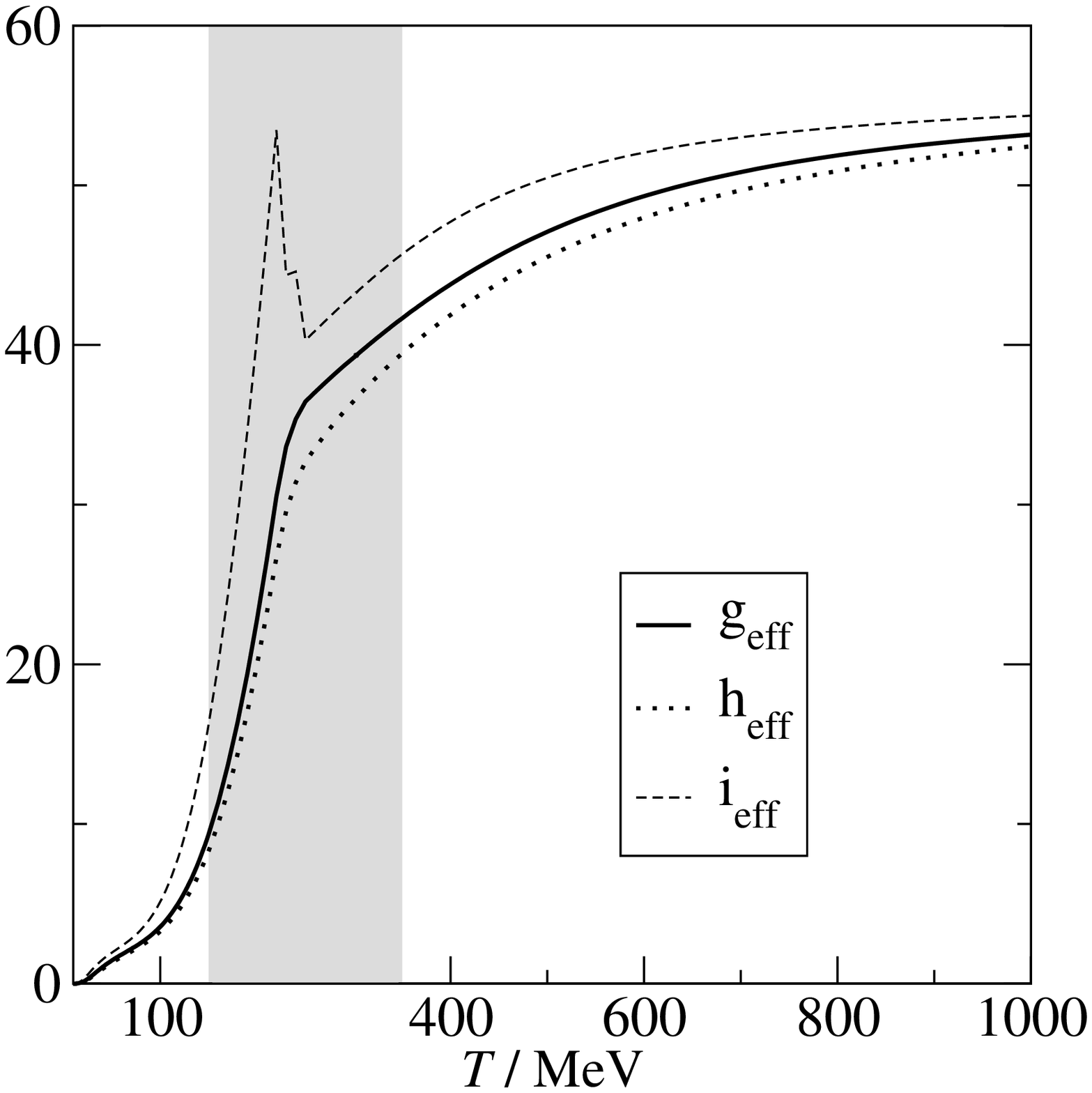}%
~~\epsfysize=5.0cm\epsfbox{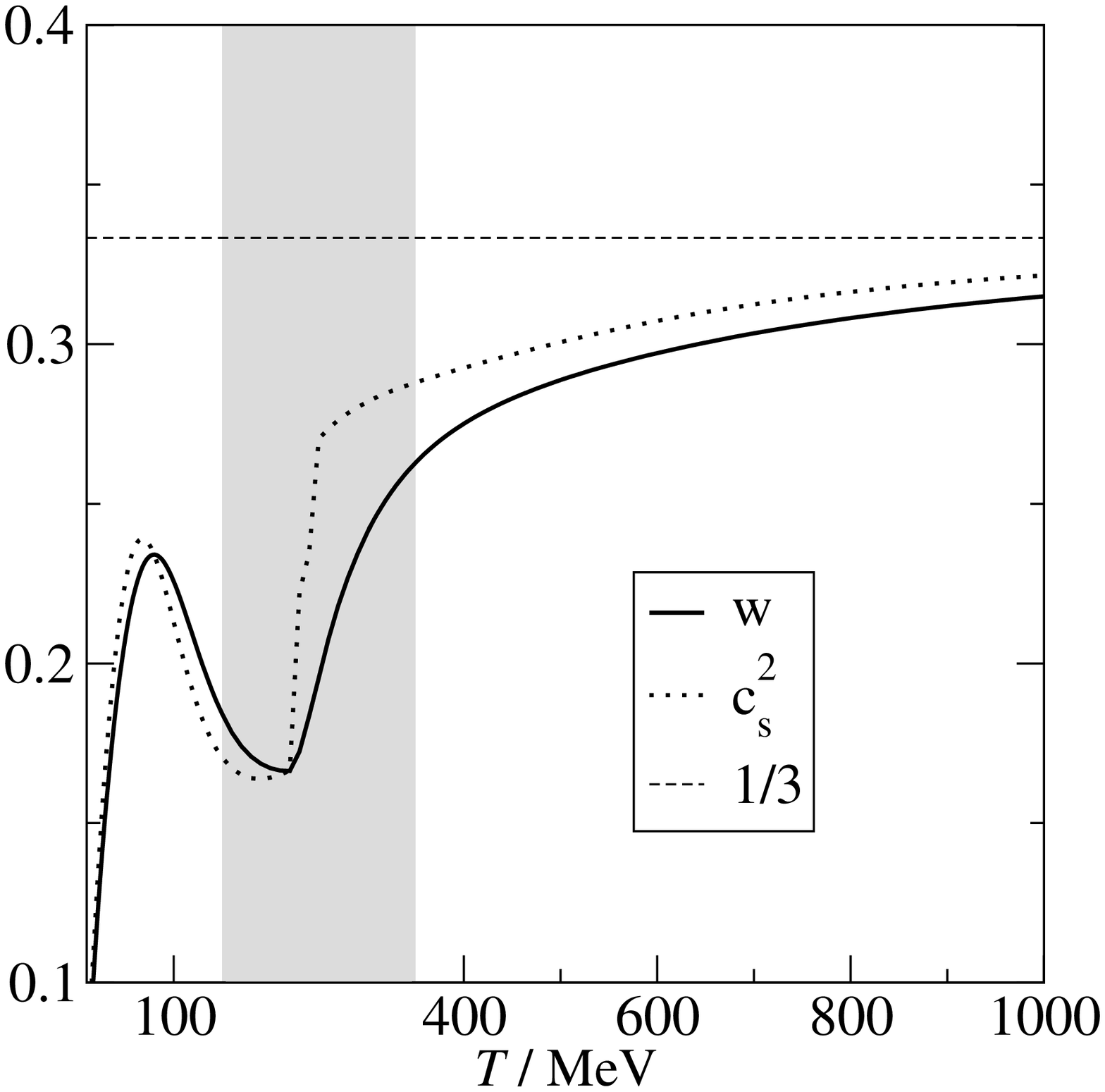}%
}

\caption[a]{%% \small
Left:  Phenomenological
interpolating curves (solid lines) for the QCD pressure at $\Nf = 3, 4$. 
The shaded interval
corresponds to the transition region where the results can 
be reliably determined with lattice simulations only.
Middle: $g_\rmi{eff}, h_\rmi{eff}, i_\rmi{eff}$ as defined 
in~\eq\nr{ieff}, for $\Nf = 4$.
Right:   the equation-of-state parameter $w$ 
and the speed of sound squared $c_s^2$, for the same system. 
} 

\la{fig:qcdeos}
\la{fig:qcdpoT4}
\end{figure}
%%%%%%%%%%%%%%%%%%%%%%%%%%%%%%%%%%%%%%%%%%%%%%%%%%%%%%%%%%%%%%%%%%%%%%%%%%%

Results for all of these quantities, based on our interpolation, 
are shown in \fig\ref{fig:qcdeos}(middle, right). 
It can be seen that quantities involving 
derivatives show a significant amount of structure around 
the QCD crossover, even if there were no singularities. 
We remark that to smooth the behaviour 
we have evaluated $p(T,\boldmu)$ with a relatively 
sparse temperature grid in the critical region.

Clearly, it is important to correct the results 
in the ``shaded region'' by using results from future lattice 
simulations of the type in Refs.~\cite{Nfcp}--\cite{Nfbi}.
In particular, the recent Refs.~\cite{Nfwu,Nfbi} display
direct results for $c_s^2$ and $w$, respectively. Unfortunately, it does not 
appear that these results would be useful for our present purposes: 
they for instance fail to reproduce the significant rise 
in $w$ and $c_s^2$ that is seen in \fig\ref{fig:qcdeos}(right)
at temperatures down from 
the critical one, displaying rather a much deeper dip (down to $\sim 0.1$)
around the critical region, and then rising at most slightly as the 
temperature is lowered. Therefore, it could be feared that the dip itself 
is affected by the unphysically heavy quark masses 
that are used in the simulations.  

We finally comment on the peak visible in $i_\rmi{eff}$
in \fig\ref{fig:qcdeos}(middle). While the details are of course
not captured by our phenomenological recipe, 
the fact that a peak exists in the 
heat capacity is not unexpected for rapid crossovers;
in second order phase transition, 
the heat capacity even diverges as $T \to \Tc$. 

%%%%%%%%%%%%%%%%%%%%%%%%%%% SECTION %%%%%%%%%%%%%%%%%%%%%%%%%%%%%%%%%%%%%
%
\section{Phenomenological results for the Standard Model}
\la{se:phenew}

While in heavy ion collisions at most strongly interacting
particles have time to thermalise, the expansion rate is much
smaller in cosmology, so that all Standard Model degrees 
of freedom do reach thermal equilibrium, and remain thermalised
until neutrino decoupling at around $T \sim $~MeV. Therefore, 
their contributions need to be added to the QCD pressure. 
In practice, we count gluons and the four lightest quarks
as the QCD degrees of freedom, while the bottom and top quark 
are treated as part of the ``weakly interacting'' sector, so 
that the result splits into a sum of two terms.

We will assume that it is sufficient to treat the weakly
interacting sector at 1-loop level. That is, we construct
the free energy density $f$ in the presence of a Higgs 
expectation value $v$, temperature $T$, and chemical 
potentials $\mu_i$, according to  
\ba
 f(v,T,\boldmu) = -\fr12 \nu^2(\bmu) v^2 + \fr14 \lambda(\bmu) v^4 
 + \sum_i \sigma_i \mathcal{J}_i(m_i(v),T,\mu_i)
 \;, \la{f1l}
\ea
where the sum extends over all physical degrees of freedom,
with their proper degeneracies; $\sigma_i = +1$ ($-1$) for 
bosons (fermions); and the tree-level masses $m_i(v)$
depend on $v$ in the standard way (it is sufficient at this 
order to work in unitary gauge). 
For scalar ($\mathcal{J}_\rmi{s}$),
vectors ($\mathcal{J}_\rmi{v}$) 
and fermions ($\mathcal{J}_\rmi{f}$), 
\ba
 \mathcal{J}_\rmi{s} & = & 
 -\frac{m^4}{64\pi^2}
 \biggl(
   \ln\frac{\bmu^2}{m^2} + \fr32 
 \biggr)
 + \frac{T^4}{4\pi^2}
 \int_0^\infty \! {\rm d} x \, x^{\fr12}
 \ln\Bigl(
   1 - e^{-\sqrt{x+y}} 
 \Bigr)_{y = \frac{m^2}{T^2}}
 \;, \\
 \mathcal{J}_\rmi{v} & = & 
 -\frac{m^4}{64\pi^2}
 \biggl(
   \ln\frac{\bmu^2}{m^2} + \fr56 
 \biggr)
 + \frac{T^4}{4\pi^2}
 \int_0^\infty \! {\rm d} x \, x^{\fr12}
 \ln\Bigl(
   1 - e^{-\sqrt{x+y}} 
 \Bigr)_{y = \frac{m^2}{T^2}}
 \;, \\
 \mathcal{J}_\rmi{f} & = & 
 -\frac{m^4}{64\pi^2}
 \biggl(
   \ln\frac{\bmu^2}{m^2} + \fr32 
 \biggr)
 + T^4 F_1\biggl( \frac{m^2}{T^2}, \frac{\mu}{T} \biggr)
 \;. 
\ea
The renormalised pressure is then given by
\be
  p(T,\boldmu) =  {\mbox{min}}_v f(v,0,\vec{0}) 
                 -{\mbox{min}}_v f(v,T,\boldmu) 
  \;. \la{ewpres}
\ee

%%%%%%%%%%%%%%%%%%%%%%%%%%%%%%%%% FIGURE %%%%%%%%%%%%%%%%%%%%%%%%%%%%%%%%%
\begin{figure}[t]

\centerline{%
\epsfysize=5.0cm\epsfbox{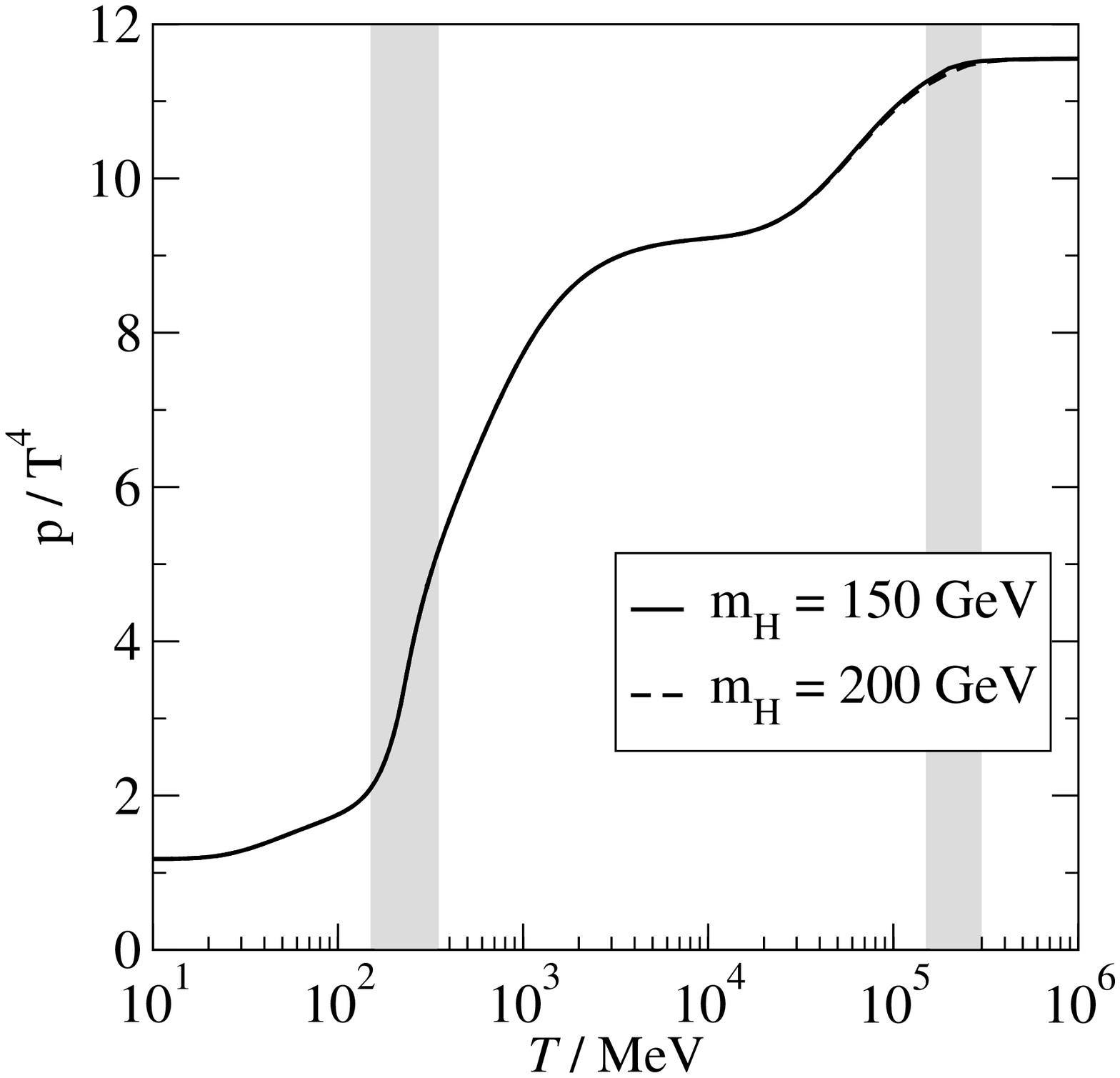}%
~~\epsfysize=5.0cm\epsfbox{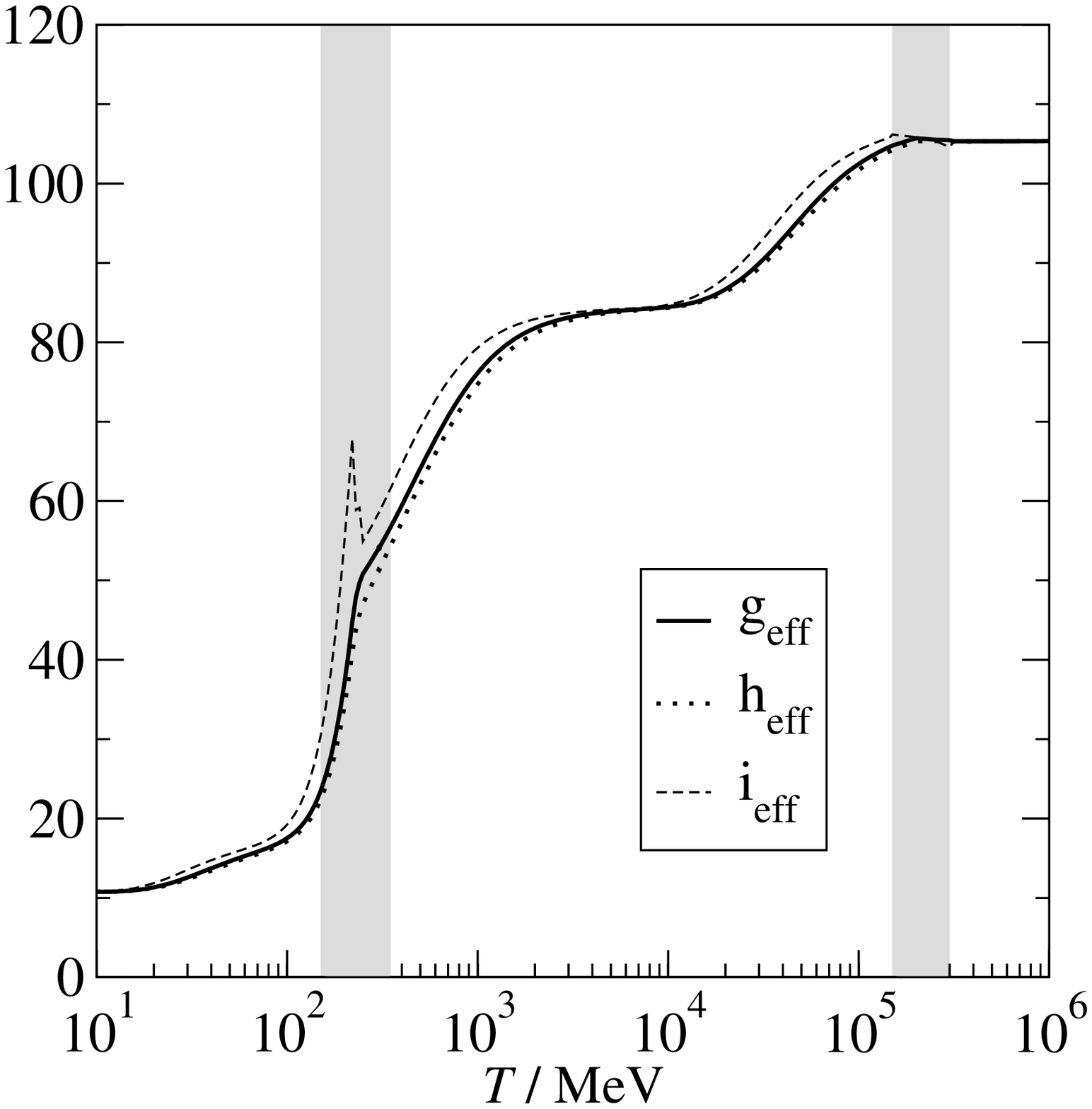}%
~~\epsfysize=5.0cm\epsfbox{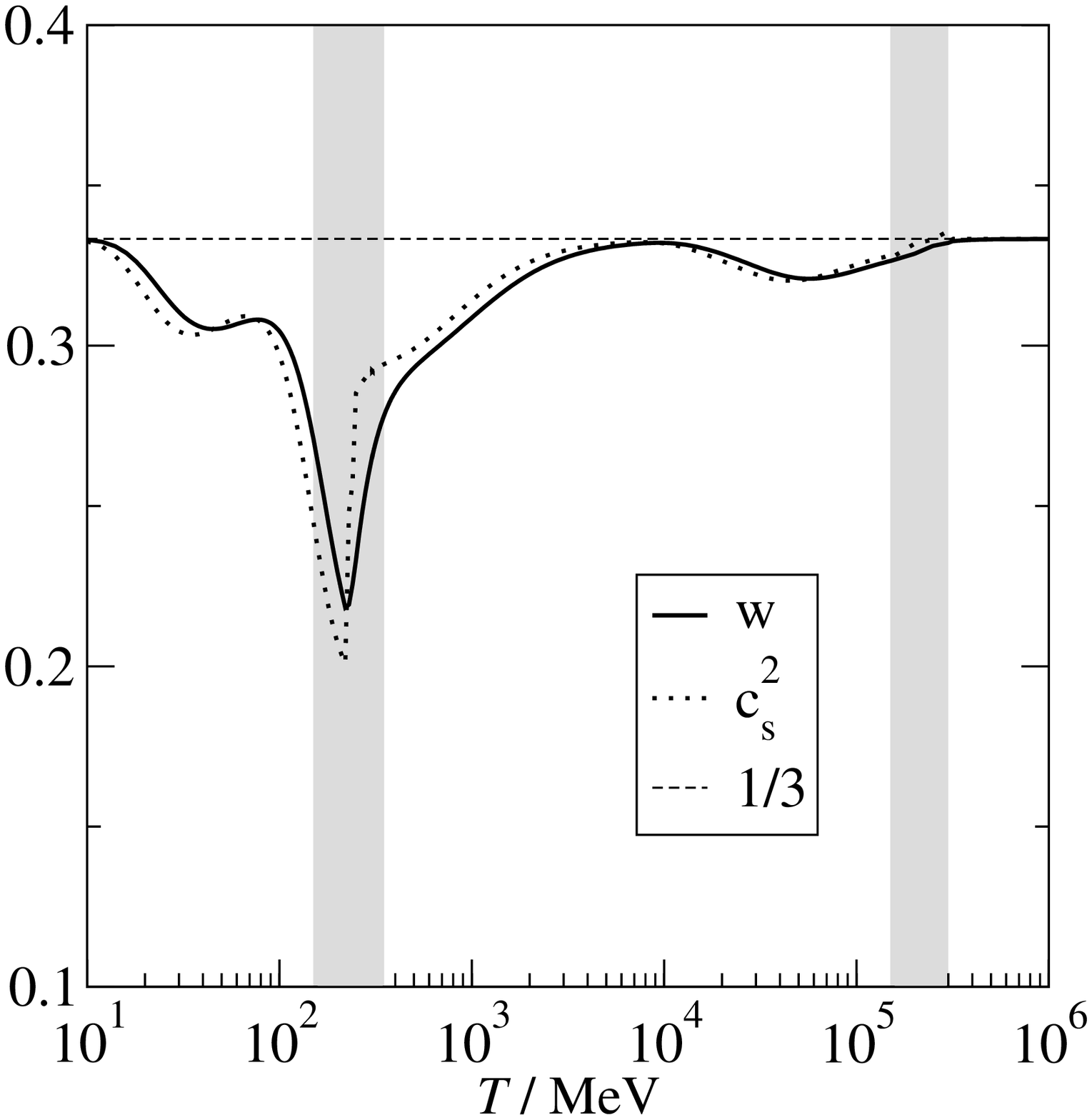}%
}

\caption[a]{%% \small
Left: The Standard Model pressure for $m_H =$ 150~GeV, 200~GeV. 
The shaded intervals correspond to the QCD and electroweak 
transition regions.
Middle: $g_\rmi{eff}, h_\rmi{eff}, i_\rmi{eff}$ as defined 
in~\eq\nr{ieff}, for $m_H =$ 150~GeV.
Right: the equation-of-state parameter $w$ 
and the speed of sound squared $c_s^2$, for the same system.
Various sources of uncertainties are discussed in the text.
} 

\la{fig:eweos}
\la{fig:ewpoT4}
\end{figure}
%%%%%%%%%%%%%%%%%%%%%%%%%%%%%%%%%%%%%%%%%%%%%%%%%%%%%%%%%%%%%%%%%%%%%%%%%%%

The renormalised pressure depends on a number of parameters
defined in the $\msbar$ scheme: 
the Higgs potential parameters $\nu^2(\bmu)$, $\lambda(\bmu)$
(cf.\ \eq\nr{f1l}); and the weak gauge and 
the top and bottom Yukawa couplings 
$g_w^2(\bmu)$, $h_t^2(\bmu)$, $h_b^2(\bmu)$
(through the tree-level masses).
The first four of these we express through the Fermi constant 
and the $W^\pm$, Higgs, and top pole masses, employing the explicit 
1-loop relations listed in Ref.~\cite{generic}, while the last one 
is fixed through the bottom mass in the $\msbar$ scheme~\cite{pdg}. 
Given that the electroweak theory contains a multitude 
of scales, both zero temperature and thermal, we simply
choose a fixed $\bmu = 100$~GeV for the weakly interacting
part of the pressure
(we have varied the scale by a factor 0.5 ... 2.0, and
seen that the dependence is invisible on our resolution).

Let us remark that \eq\nr{ewpres} suffers from the 
problem that it leads to a first order electroweak phase 
transition at a certain temperature, while there is none
if the theory is treated more carefully~\cite{isthere,su2u1}.
In practice this does not lead to any serious complications, 
however: we again smooth the behaviour by 
evaluating $p(T,\boldmu)$ with a sparse temperature 
grid around the critical region. In our figures, 
we shade the corresponding temperature interval, where 
our estimates are qualitative at best. 

With these reservations, the whole Standard Model
pressure, and the parameters defined 
in \eqs\nr{ieff}, \nr{wT}, \nr{csT}, 
are shown in \fig\ref{fig:eweos}.

At temperatures above the electroweak scale, our results
are already very close to the ideal gas results. Recently, 
higher loop corrections in this
region have been considered in some detail~\cite{gv}.
The authors find a rather more significant
deviation from the ideal gas value, 
due for instance to the top Yukawa coupling. 
We have not implemented these corrections, however, since they 
would require a correspondingly higher order computation in the broken 
symmetry phase. Though such a computation exists in principle up to 2-loop 
level~\cite{fh}, we did not consider the non-trivial
challenges posed by its numerical evaluation for general masses
to be worth tackling at present, given that in the quantities 
plotted in \fig\ref{fig:eweos}(right), the 2-loop contributions 
(which do not contribute to the trace anomaly on the 
symmetric phase side) are expected to largely cancel out. 
Nevertheless, it would be important to finalise
this computation, if physics is made with the temperature
interval $T = 10 ... 100$~GeV, where the $W^\pm, Z^0$ bosons
and top quark cross their mass thresholds. 

Finally, once a definite Higgs model is available, 
it will of course be important to carry out lattice simulations 
for the transition region. Fortunately, for the electroweak
theory this can be achieved within the dimensionally reduced
effective theory~\cite{generic}, whereby also all fermions 
with their physical Yukawa couplings can be fully accounted for. 

% \newpage

%%%%%%%%%%%%%%%%%%%%%%%%%%% SECTION %%%%%%%%%%%%%%%%%%%%%%%%%%%%%%%%%%%%%
%
\section{Conclusions}
\la{se:concl}

The functional
dependence of the QCD pressure at a high temperature $T$ 
on most of the parameters of the theory (number of colours $\Nc$, 
number of active massless flavours $\Nf$, 
and quark chemical potentials $\mu_i$) is known up to relative 
order $\rmO(g^6\ln(1/g))$, while the dependence on the quark 
masses $m_i$ has gained much less interest.
The purpose of this note has been to study whether it indeed
is justified to consider the effects of finite non-zero quark masses 
at the level of the non-interacting Stefan-Boltzmann law 
(i.e.\ $\rmO(g^0)$), as has been the standard procedure. 
For this purpose, we have determined the corrections of order 
$\rmO(g^2)$ in full generality in the $\msbar$ scheme, 
and presented a numerical evaluation of all the integrals 
that appear in this result. 

We find that while the $\rmO(g^2)$ corrections 
are in general 20...30\% (\fig\ref{fig:mass}(left)), 
they are numerically at most 5\%
for the quark mass dependence (\fig\ref{fig:mass}(right)).
This is perhaps in accord with previous observations
according to which quarks are fairly perturbative as soon as they
are deconfined, even though gluons do display strong interactions
up to very high temperatures. 

Finally, we have sketched
educated ``guesses'' for the thermodynamic quantities that play a role
in various physical contexts, for temperatures between the QCD and
electroweak scales. 
For the case of heavy ion collisions, 
in particular, it is
perhaps relevant to keep in mind that if the charm quark does thermalise, 
it has a rather significant effect even at relatively low temperatures
(\fig\ref{fig:qcdpoT4}(left)). Of course, it is by no means clear
whether such a thermalization should take place in practice~\cite{mt}.

In order to improve on our QCD results, the missing perturbative
$\rmO(g^6)$ computations and, naturally, lattice simulations in 
the transition region, with physical values of the quark masses, 
remain to be completed. 

For the full Standard Model, we have presented similar
guesses for the various quantities that are 
relevant for expansion rate and particle decoupling computations 
(Figs.~\ref{fig:ewpoT4}).
Although the deviations from previous phenomenological estimates
that have appeared in the literature~\cite{olddm2,dm2} 
are in general fairly small, we nevertheless hope that our results 
help for their part to 
gauge the systematic uncertainties that still exist in these quantities.  

In particular, we have stressed the need for repeating 
the computations of Ref.~\cite{gv} in the broken symmetry phase and, 
of course, the need for effective theory lattice simulations in the 
transition region, once the electroweak model / Higgs mass is known.

%%%%%%%%%%%%%%%%%%%%%%%%%%%%%%%%%%%%%%%%%%%%%%%%%%%%%%%%%%%%%%%%%%%%
%
\section*{Acknowledgements}

We acknowledge useful discussions with 
P.~Huovinen, 
K.~Kajantie, 
K.~Rummukainen,  
M.~Shaposhnikov, and
A.~Vuorinen, and thank J.~Engels for providing
lattice data from Ref.~\cite{bi1}.

% \newpage

%-------------------------------------------------------------------

\appendix
%% \renewcommand{\thesection}{Appendix~\Alph{section}}
%% \renewcommand{\thesubsection}{\Alph{section}.\arabic{subsection}}
%% \renewcommand{\theequation}{\Alph{section}.\arabic{equation}}

%%%%%%%%%%%%%%%%%%%%%%%%%%%%%%%%%%%%%%%%%%%%%%%%%%%%%%%%%%%%%%%%%%%%%%%%

%%%%%%%%%%%%%%%%%%%%%%%%% SECTION %%%%%%%%%%%%%%%%%%%%%%%%%%%%%%%%%%%%%
%
\section{Outline of the computation}

In this Appendix we present a few details for the computation
leading to \eqs\nr{aE1}--\nr{aE7}. We concentrate on the 
fermionic contributions; the bosonic ones are elementary. 

We write the fermionic contributions in terms of the renormalised 
gauge coupling $\g^2$ and the renormalised quark masses $\mi$, 
$i = 1,...,\Nf$. The master integrals emerging are 
\ba
 I_\rmi{b} & \equiv & \Tint{P_\rmi{b}} \frac{1}{P_\rmi{b}^2} 
 \;, \la{Ibdef} \\
 J_\rmi{f}(m,\mu) & \equiv & 
 \fr12 \Tint{P_\rmi{f}} \ln (\tilde  P_\rmi{f}^2 + m^2) 
 \;, \la{Jfdef} \\
 I_\rmi{f}(m,\mu) & \equiv & \Tint{P_\rmi{f}} 
 \frac{1}{\tilde P_\rmi{f}^2 + m^2} 
 \;, \la{Ifdef} \\ 
 H_\rmi{f}(m,\mu) & \equiv & \Tint{P_\rmi{f},Q_\rmi{f}} 
 \frac{1}{(\tilde P_\rmi{f}^2 + m^2)
          (\tilde Q_\rmi{f}^2 + m^2) 
          (\tilde P_\rmi{f} - \tilde Q_\rmi{f})^2}
 \;, \la{Hfdef}
\ea
where $P_\rmi{b}$, $P_\rmi{f}$ denote bosonic and 
fermionic Matsubara four-momenta, respectively, and 
$\tilde P_\rmi{f} \equiv P_\rmi{f} + (-i\mu,\vec{0})$
includes the chemical potential $\mu$.
With this notation, the fermionic contributions to the 
parameters of interest are
\ba
 T^4 \aE{1}^\rmi{f} & = & 4 C_A \sum_{i = 1}^{\Nf}
 J_\rmi{f}(\mi,\mu_i) \;, \la{aE1f} \\
%%%%%%%%%%%%%%%%%%%%%%%%%%%%%%%%%%%%%%%%%%%%%%%%%%%%%
 T^4 \aE{2}^\rmi{f}  & = &
 2 C_A \hat \delta_1 m^2 \sum_{i=1}^{\Nf} \mi^2 I_\rmi{f}(\mi,\mu_i)
 + \nn & + & 
%%%%%%%%%%%%%%%%%%%%%%%
 d_A \sum_{i=1}^{\Nf}
 \biggl\{ 
%   \frac{1-d}{2}
%   \Bigl[
%     I_\rmi{f}^2(\mi,\mu_i) 
%     - 2 I_\rmi{b} I_\rmi{f}(\mi,\mu_i)  
%   \Bigr]
   \frac{d-1}{2}
   \Bigl[ 2 I_\rmi{b}
     - I_\rmi{f}(\mi,\mu_i) 
   \Bigr] I_\rmi{f}(\mi,\mu_i)  
  + 2 \mi^2 H_\rmi{f}(\mi,\mu_i)
 \biggr\}
 \;, \\
%%%%%%%%%%%%%%%%%%%%%%%%%%%%%%%%%%%%%%%%%%%%%%%%%%%%%
 \frac{1}{(4\pi)^2}\aE{7}^\rmi{f}  & = &
 % \Bigl[ 
   \hat \delta_1 g^2 + \fr23 \sum_{i=1}^{\Nf}
  \frac{{\rm d} I_\rmi{f}(\mi,\mu_i)}{{\rm d} \mi^2}
 % \Bigr]
 \;, \la{aE7f} 
\ea
where $\hat \delta_1 m^2$, $\hat \delta_1 g^2$ are counterterms
defined by writing the bare mass parameter and gauge coupling as 
$
 m_{Bi}^2 = \mi^2(1 + g^2 \hat \delta_1 m^2)
$, 
$
 g_B^2 = g^2(1 + g^2 \hat \delta_1 g^2)
$;
it is understood that only the fermionic part of 
$\hat \delta_1 g^2$ is considered; and $d \equiv 3 - 2 \epsilon$.

The next step is the evaluation of the Matsubara sums appearing
in the master integrals.
For the 1-loop structures
(\eqs\nr{Ibdef}--\nr{Ifdef}), it is straightforward to obtain
\ba
 I_\rmi{b} & = & \int\! \frac{{\rm d}^d \vec{p}}{(2\pi)^d}
 \frac{1}{2|\vec{p}|}\Bigl[ 1 + 2 \nB(|\vec{p}|)\Bigr]
 \;,  \la{Ib} \\  
 J_\rmi{f}(m,\mu) & = & \int\! \frac{{\rm d}^d \vec{p}}{(2\pi)^d}
 \frac{-\vec{p}^2}{2dE}\Bigl[ 1 - \nF(E-\mu) - \nF(E+\mu)
 \Bigr]_{E=\sqrt{\vec{p}^2 + m^2}}
 \;,  \la{Jf} \\  
 I_\rmi{f}(m,\mu) & = & \int\! \frac{{\rm d}^d \vec{p}}{(2\pi)^d}
 \frac{1}{2E}\Bigl[ 1 - \nF(E-\mu) - \nF(E+\mu)
 \Bigr]_{E=\sqrt{\vec{p}^2 + m^2}}
 \;,    \la{If}
\ea
where we have carried out a partial integration 
after the sum in $J_\rmi{f}$,  and 
\be
 \nB(E) \equiv \frac{1}{e^{\beta E} - 1}
 \;, \quad
 \nF(E) \equiv \frac{1}{e^{\beta E} + 1}
 \;.
\ee
As is well known, the momentum integral 
in \eq\nr{Ib} can be carried out explicitly, 
$
  I_\rmi{b} = \pi^{d/2} T^{d}
  \Gamma(1-d/2)\zeta(2 - d)/2\pi^2 T
$,
but the ones in \eqs\nr{Jf}, \nr{If} with $m,\mu \neq 0$
cannot in general be integrated in closed form.

For the genuine 2-loop integral $H_\rmi{f}$ in \eq\nr{Hfdef} 
the sums are slightly more complicated, so we give here some details.
The method we employ follows 
the standard procedure~\cite{jk} (see also Refs.~\cite{old,ll}). 
The twofold sum over the Matsubara modes 
is first written as a threefold sum with a Kronecker delta-function, 
and the delta-function is then written as 
$\delta(p_0) = T \int_0^\beta \! {\rm d} x \, \exp({i p_0 x})$.
The sums can now be performed: 
\ba
 T\sum_{p_\rmi{b}} \frac{e^{i p_\rmi{b} x}}{p_\rmi{b}^2 + E^2} & = &  
 \frac{1}{2E} n_\rmi{B}(E) \Bigl[ 
 e^{(\beta-x)E} + e^{x E}
 \Bigr]  
 \;, \\
 T\sum_{p_\rmi{f}} \frac{e^{i \tilde p_\rmi{f} x}}{\tilde p_\rmi{f}^2 + E^2} 
 & = & 
 \frac{1}{2E} \Bigl[ n_\rmi{F}(E+\mu)
 e^{(\beta-x)E + \beta\mu} - n_\rmi{F}(E-\mu) e^{x E}
 \Bigr]  
 \;,
\ea
where $p_\rmi{b}$, $p_\rmi{f}$ denote the bosonic 
and fermionic Matsubara frequencies, respectively; 
and $\tilde p_\rmi{f} = p_\rmi{f} - i \mu$.
The integral over $x$ is then simple.
All the exponents appearing can be written in terms of inverses
of the distribution functions $\nB$, $\nF$, and multiplying
with their explicit appearances, we are left with at most
quadratic products of the distribution functions, and fractions
containing the three ``energies''.  

The fractions containing the energies can be organized in a 
transparent form, once we introduce the zero-temperature objects
\ba 
 H_\rmi{vac}(m_1^2,m_2^2,m_3^2) 
 & = & \int \! \frac{{\rm d}^{4-2\epsilon}P}{(2\pi)^{4-2\epsilon}}
       \int \! \frac{{\rm d}^{4-2\epsilon}Q}{(2\pi)^{4-2\epsilon}}
       \frac{1}{[P^2+m_1^2][Q^2+m_2^2][(P-Q)^2+m_3^2]}
 \;, \hspace*{1cm} \la{Hvacdef} \\
 \Pi(Q^2;m_1^2,m_2^2) & = & 
 \int \! \frac{{\rm d}^{4-2\epsilon} P }{(2\pi)^{4-2\epsilon}}
 \frac{1}{[P^2+m_1^2][(P-Q)^2 + m_2^2]}
 \;, \la{B0} \\
 \Delta(Q^2;m_3^2) & = & \frac{1}{Q^2 + m_3^2}
 \;. 
\ea
Indeed, carrying out the integrals over $P_0, Q_0$ in these functions, 
one obtains similar energy fractions. Making furthermore use 
of the O($4-2\epsilon$)
rotational invariance of the $Q$-dependence in \eq\nr{B0}, 
which is present once also the 
integration over $\vec{p}$ is performed, 
the various fractions can be identified with each other. 

In order to write the subsequent result in a compact but generic form, 
we introduce the notation
\be
 E_i \equiv \sqrt{\vec{p}_i^2 + m_i^2}
 \;, \quad
 P_i \equiv (E_i,\vec{p}_i)
 \;, \quad
 P_i \cdot P_j \equiv E_i E_j - \vec{p}_i\cdot \vec{p}_j
 \;,  
\ee
and denote
\be
 n_\pm(E_i) \equiv \left\{
 \begin{array}[c]{ll}
  \nB(E_i) & \mbox{for bosons}\; (\equiv E_3) \\ 
  -\nF(E_i\pm\mu) & \mbox{for fermions}\; (\equiv E_1,E_2)
 \end{array}
 \right. 
 \;.
\ee
Then, allowing for generality for three different masses,  
like in \eq\nr{Hvacdef}, 
\ba
 H_\rmi{f} & = &  
 H_\rmi{vac}(m_1^2,m_2^2,m_3^2)  + 
 \nn & + & 
 \sum_{i\neq j\neq k} \sum_{\sigma=\pm 1}
 \int \! \frac{{\rm d}^d \vec{p}_i}{(2\pi)^d}
 \frac{n_\sigma(E_i)}{2 E_i} \Pi(-m_i^2;m_j^2,m_k^2) + 
 \nn & + & 
 \sum_{i\neq j\neq k} \sum_{\sigma,\tau=\pm 1}
 \int \! \frac{{\rm d}^d \vec{p}_i}{(2\pi)^d}
 \int \! \frac{{\rm d}^d \vec{p}_j}{(2\pi)^d}
 \frac{n_\sigma(E_i) n_\tau(E_j)}{4 E_i E_j} 
 \Delta[-(\sigma P_i - \tau P_j)^2;m_k^2]
 \;, \la{genH}
\ea
where 
$\sum_{i\neq j\neq k}\equiv \sum_{(i,j,k)=(1,2,3),(2,3,1),(3,1,2)}$.
Individual terms in this sum may contain infrared poles (or, after 
performing some of the integrations in complex plane, imaginary parts),
but the expression as a whole is finite and real for $\epsilon\neq 0$.

We return now to the case of physical interest 
($m_1^2=m_2^2\equiv m^2; m_3^2\equiv 0$), and ignore 
all temperature-independent terms. We note, 
furthermore, that the contribution originating from 
the last term in \eq\nr{genH}
for $(i,j,k)=(3,1,2)$, contains
$ 
 \Delta[-(P_3+P_1)^2;m^2] + \Delta[-(P_3-P_1)^2;m^2]
$
which vanishes, given that $P_1^2 + P_3^2 = m^2$.
The same is true for $(i,j,k)=(2,3,1)$.
This leaves us with
\ba
 H_\rmi{f}(m,\mu) \!\! & = & \!\! [\mbox{temperature-independent terms}] + 
 \nonumber \\[2mm] & + & \!\!
 I_\rmi{b}\, \Pi(0;m^2,m^2) + 2 I_\rmi{f}(m;\mu) \, \Pi(-m^2;m^2,0) + 
 \nonumber \\[2mm] & + & \!\!\!
 \int \! \frac{{\rm d}^d \vec{p}_1}{(2\pi)^d}
 \int \! \frac{{\rm d}^d \vec{p}_2}{(2\pi)^d}
 \frac{n_-(E_1)n_+(E_2)+n_+(E_1)n_-(E_2)}{8 E_1 E_2}
 \frac{1}{\vec{p}_1\cdot \vec{p}_2 -m^2 - E_1 E_2 } + 
 \nonumber \\[2mm] & + & \!\!\!
 \int \! \frac{{\rm d}^d \vec{p}_1}{(2\pi)^d}
 \int \! \frac{{\rm d}^d \vec{p}_2}{(2\pi)^d}
 \frac{n_-(E_1)n_-(E_2)+n_+(E_1)n_+(E_2)}{8 E_1 E_2}
 \frac{1}{\vec{p}_1\cdot \vec{p}_2 -m^2 + E_1 E_2 } 
 \;, \hspace*{1.2cm} \la{Hffinal}
% \nn  
\ea
where we substituted $\vec{p}_1 \to - \vec{p}_1$ in the last term.
We can still perform the integration over 
$z\equiv \vec{p}_1\cdot\vec{p}_2/|\vec{p}_1||\vec{p}_2|$, leaving 
a rapidly convergent integral over $|\vec{p}_1|$, $|\vec{p}_2|$. 

The final step is the expansion in $\epsilon$.
The only temperature-dependent ultraviolet divergences are in 
the factorised terms on the second row in \eq\nr{Hffinal}. 
Adding together with the contributions from 
the other master integrals, as specified in \eqs\nr{aE1f}--\nr{aE7f}, 
a straightforward computation reproduces 
the fermionic parts of \eqs\nr{aE1}--\nr{aE7}.

% \newpage

%%%%%%%%%%%%%%%%%%%%%%%%% SECTION %%%%%%%%%%%%%%%%%%%%%%%%%%%%%%%%%%%%%
%
\section{Functions determining the mass dependence}

The functions that appear in \eqs\nr{aE1}--\nr{aE7} are defined as
\ba
 F_1(y,\hat\mu) \!\! & \equiv & \!\! 
 \frac{1}{24\pi^2} \int_0^\infty \! {\rm d} x\, 
 \biggl[ \frac{x}{x+y} \biggr]^{\fr12}
 \Bigl[
   \nFm{x} + \nFp{x} 
 \Bigr] \, x
 \; , \la{F1def} \\ 
 F_2(y,\hat\mu) \!\! & \equiv & \!\! 
 \frac{1}{8\pi^2} \int_0^\infty \! {\rm d} x\, 
 \biggl[ \frac{x}{x+y} \biggr]^{\fr12}
 \Bigl[
   \nFm{x} + \nFp{x} 
 \Bigr] 
 \; , \\ 
 F_3(y,\hat\mu) \!\! & \equiv & \!\! 
 - \int_0^\infty \! {\rm d} x\, 
 \biggl[ \frac{x}{x+y} \biggr]^{\fr12}
 \Bigl[
   \nFm{x} + \nFp{x} 
 \Bigr] \frac{1}{x} 
 \; ,  \\ 
 F_4(y,\hat\mu) \!\! & \equiv & \!\! 
 \frac{1}{(4\pi)^4} 
 \int_0^\infty \! {\rm d} x_1\, \int_0^\infty \! {\rm d} x_2\, 
 \frac{1}{\sqrt{x_1 + y}\sqrt{x_2 + y}} \times 
 \nn & & \!\!
 \times\biggl\{ 
 \Bigl[ \nFm{x_1}\nFp{x_2} + \nFp{x_1} \nFm{x_2} \Bigr] 
 \times
 \nn & & \times \ln\biggl[
 \frac{\sqrt{x_1 + y}\sqrt{x_2 +y} + y - \sqrt{x_1 x_2}}
      {\sqrt{x_1 + y}\sqrt{x_2 +y} + y + \sqrt{x_1 x_2}}\biggr] + 
 \nn & & \!\! + 
 \Bigl[ \nFm{x_1}\nFm{x_2} + \nFp{x_1} \nFp{x_2} \Bigr] 
 \times
 \nn & & \times \ln\biggl[
 \frac{\sqrt{x_1 + y}\sqrt{x_2 +y} - y + \sqrt{x_1 x_2}}
      {\sqrt{x_1 + y}\sqrt{x_2 +y} - y - \sqrt{x_1 x_2}}\biggr]
 \biggr\} 
 \;,  \la{F4def}
\ea
where 
\be
 \hat \nF(x) \equiv \frac{1}{e^x + 1}
 \;.
\ee
These functions are related to the functions $J_\rmi{f}$,
$I_\rmi{f}$ and $H_\rmi{f}$ defined in \eqs\nr{Jfdef}--\nr{Hfdef}: 
the medium-modified part of $J_\rmi{f}$ reads $T^4 F_1$ for $\epsilon = 0$;
the medium-modified part of $I_\rmi{f}$ reads $-T^2 F_2$ for $\epsilon = 0$;
the medium-modified part of ${\rm d}I_\rmi{f}/{\rm d} m^2$ 
reads $-F_3/(4\pi)^2$ for $\epsilon = 0$; 
and the ``non-factorizable'' part
of $H_\rmi{f}$ (the last two terms in \eq\nr{Hffinal})
reads $T^2 F_4$ for $\epsilon = 0$.
The functions $F_1, F_2, F_3$ are related by 
\be
 F_2(y,\hat\mu) = -2 \frac{\partial F_1(y,\hat\mu) }{ \partial y}
 \;, \quad
 F_3(y,\hat\mu) = (4\pi)^2 \frac{ \partial F_2(y,\hat\mu) }{ \partial y}
 \;.
\ee

The functions introduced possess some solvable
limiting values. For $y\to 0$,
\ba
 F_1(0,\hat\mu) & = & \frac{7\pi^2}{720} + \frac{\hat\mu^2}{24} + 
 \frac{\hat\mu^4}{48 \pi^2} 
 \;, \la{F1T0} \\
 F_2(0,\hat\mu) & = & \frac{1}{24} + \frac{\hat\mu^2}{8\pi^2} 
 \;, \\ 
 F_3(0,\hat\mu) & \approx &
 \ln \frac{y}{\pi^2} + 2 \gamma_E + 
 \mathcal{D}\Bigl( \frac{\hat\mu}{\pi}\Bigr)  
 = \ln\frac{y}{16\pi^2} - 
 \Bigl[ 
  \psi\Bigl( \fr12 + i \frac{\hat\mu}{2\pi }
  \Bigr) + 
  \psi\Bigl( \fr12 - i \frac{\hat\mu}{2\pi }
  \Bigr)
 \Bigr] \la{F3T0}
 \;,  \hspace*{1cm} 
\ea
where ``$\approx$'' denotes that the logarithmic divergence 
displayed on the right-hand side needs to be subtracted
before setting $y\to 0$, and the function $\mathcal{D}$,
which has the property $\mathcal{D}(0)=0$,
corresponds to the notation in Ref.~\cite{mu}. 
The analytic expression in terms 
of $\psi(z) = \Gamma'(z)/\Gamma(z)$ 
comes from Ref.~\cite{av2}. 
For $\hat\mu\neq 0$, the function $F_4$ diverges
logarithmically at small $y$, but as it is always multiplied by $y$, 
this behaviour has little interest in the present context.
Inserting the values in \eqs\nr{F1T0}--\nr{F3T0} 
into our expressions for $\aEms{1}$, $\aEms{2}$, $\aEms{7}$, 
\eqs(3.14), (3.15), (3.20) of Ref.~\cite{av2} are reproduced.

For $y\to\infty$ and $\hat\mu$ fixed, 
the functions $F_1,F_2,F_3$ vanish 
as $\exp(-\sqrt{y})$, the function $F_4$ as 
$\exp(-2\sqrt{y})$, modulo a powerlike prefactor.
An interesting limit is obtained, however, by setting 
$y,\hat\mu$ simultaneously to infinity but keeping 
the ratio $z\equiv y/\hat\mu^2 = m^2/\mu^2$ fixed. 
This corresponds to setting the temperature to zero
but keeping $m,\mu$ finite. 
Then 
\ba
 \lim_{T\to 0} T^4  F_1\Bigl(\frac{m^2}{T^2},\frac{\mu}{T}\Bigr) & = & 
 \theta(1-z) \frac{\mu^4}{96\pi^2} 
 (2 w^3 - 3 z f_2)
 \;, \\
  \lim_{T\to 0} T^2 F_2\Bigl(\frac{m^2}{T^2},\frac{\mu}{T}\Bigr) & = & 
 \theta(1-z) \frac{\mu^2}{8\pi^2} f_2
 \;, \\ 
  \lim_{T\to 0}     F_3\Bigl(\frac{m^2}{T^2},\frac{\mu}{T}\Bigr) & = &
 \theta(1-z) \frac{2}{z} (f_2 - w) 
 \;, \\  
  \lim_{T\to 0} T^2 F_4\Bigl(\frac{m^2}{T^2},\frac{\mu}{T}\Bigr) & = & 
 \theta(1-z) \frac{\mu^2}{64\pi^4 z} (w^4 - f_2^2)
 \;,   
 \ea
where 
\be
 w \equiv \sqrt{1-z}
 \;, \quad
 f_2 \equiv \sqrt{1-z} - z \ln\frac{1+ \sqrt{1-z}}{\sqrt{z}}
 \;. 
\ee
Inserting into our expressions for $\aEms{1}$, $\aEms{2}$, 
\eqs(1) and (4) of Ref.~\cite{fr} are reproduced. 

%%%%%%%%%%%%%%%%%%%%%%%%%%%%%%%%% FIGURE %%%%%%%%%%%%%%%%%%%%%%%%%%%%%%%%%
\begin{figure}[t]

\centerline{%
\epsfysize=6.0cm\epsfbox{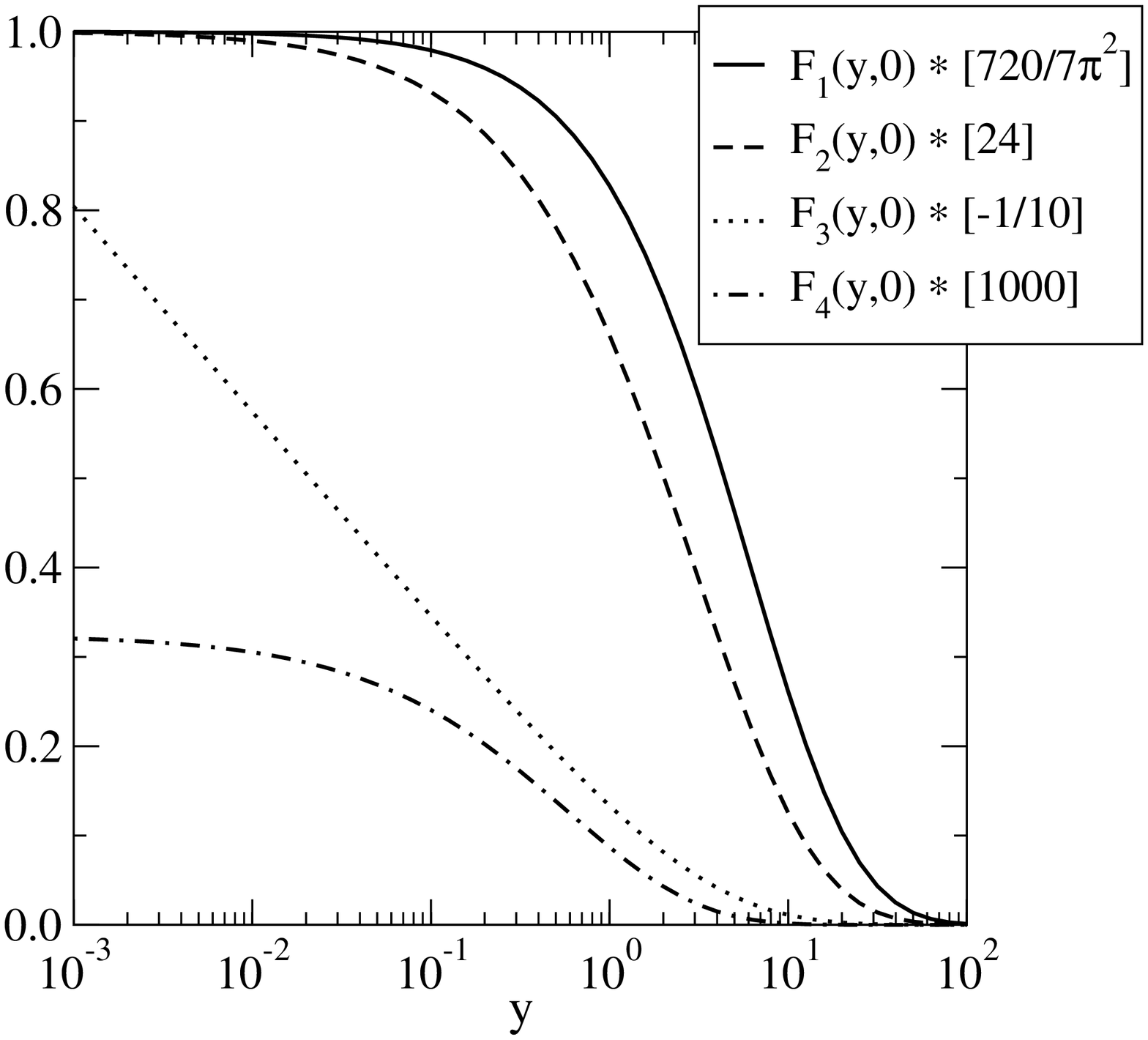}%
~~~~~\epsfysize=6.0cm\epsfbox{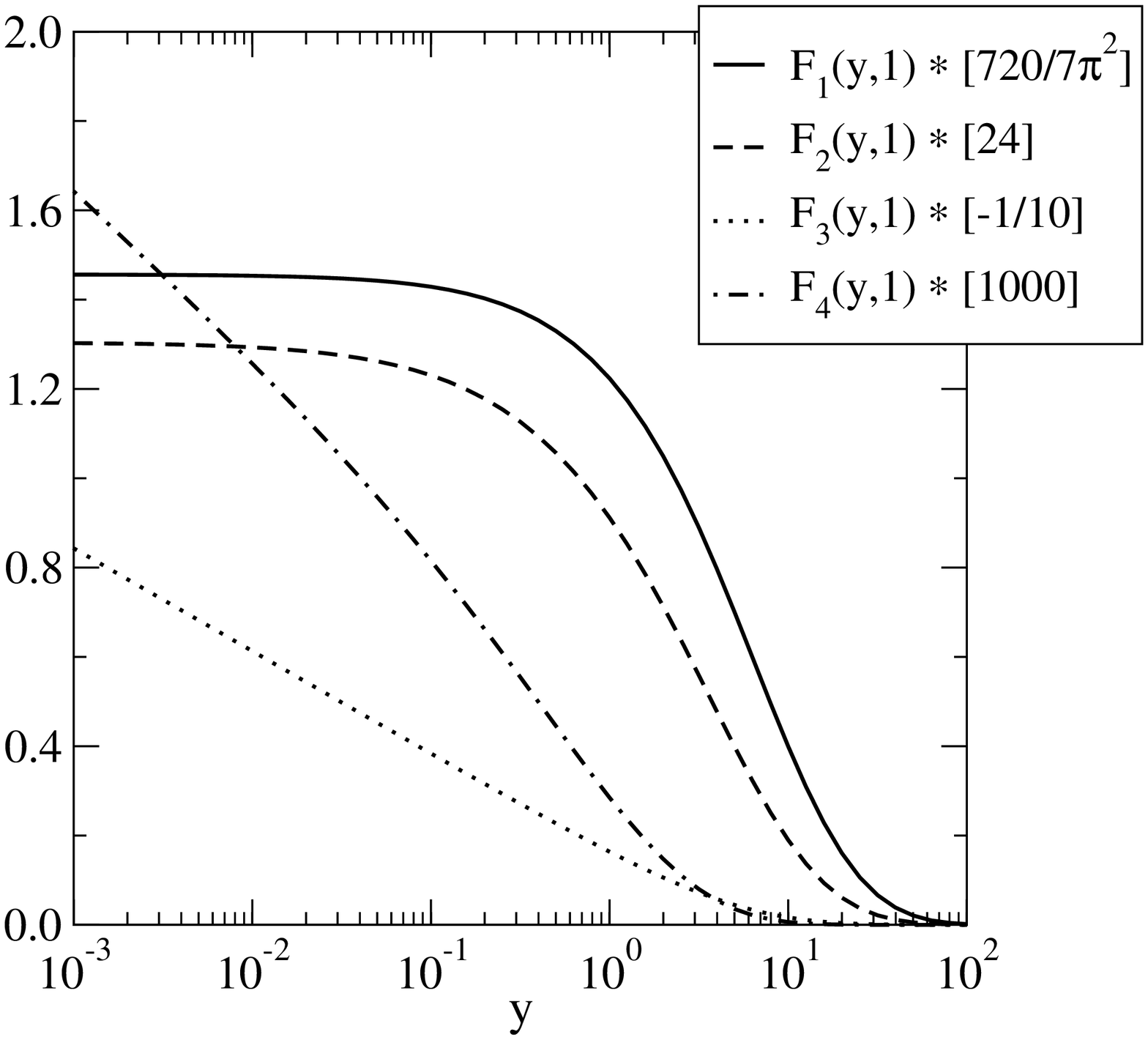}%
% ~~\epsfysize=5.0cm\epsfbox{}%
}

\caption[a]{%% \small 
The functions defined in 
\eqs\nr{F1def}--\nr{F4def}, for $\hat\mu = 0.0$ (left)
and $\hat\mu = 1.0$ (right) (all functions are even in $\hat\mu$). 
Note that the ranges of  the vertical axes 
are different in the two plots.
}

\la{fig:Fis}
\end{figure}
%%%%%%%%%%%%%%%%%%%%%%%%%%%%%%%%%%%%%%%%%%%%%%%%%%%%%%%%%%%%%%%%%%%%%%%%%%%

Unfortunately the limit $\hat\mu \to 0$,
of most interest to us in this paper, does not render any of 
the functions analytically solvable, as far as we know. 
We show the results of numerical evaluations in \fig\ref{fig:Fis}.

% \newpage

%%%%%%%%%%%%%%%%%%%%%%%%% BIBLIO %% REFERENCES %%%%%%%%%%%%%%%%%%%%%%%%%%%%%%

\end{document}